\documentclass[superscriptaddress,onecolumn,showpacs]{revtex4-2}
\usepackage[dvipdfmx]{graphicx}
\usepackage{color}
\usepackage{dcolumn}
\usepackage{amsmath}
\usepackage{lipsum}
\usepackage{dsfont}
\usepackage{epsfig}
\usepackage{soul}
\usepackage{bm}
\usepackage{amssymb}
\usepackage[dvipsnames]{xcolor}
\newcommand{\be}{\begin{equation}}
\newcommand{\ee}{\end{equation}}

\begin{document}
	\title{A theoretical framework for physically-realizable kagome metamaterials and its implications on dualities and topological edge modes}  

\author{Weijian Jiao}
\email{wjiao@tongji.edu.cn}
\affiliation{School of Aerospace Engineering and Applied Mechanics, Tongji University, Shanghai 200092, China}
\affiliation{Shanghai Institute of Aircraft Mechanics and Control,  Shanghai, 200092, China}
\affiliation{Department of Mechanical Engineering and Applied Mechanics, University of Pennsylvania, Philadelphia, Pennsylvania 19104, USA}

\author{Hang Shu}
\affiliation{Department of Mechanical Engineering and Applied Mechanics, University of Pennsylvania, Philadelphia, Pennsylvania 19104, USA}

\author{Vincent Tournat}
\affiliation{Laboratoire d'Acoustique de l'Université du Mans (LAUM), UMR 6613, Institut d'Acoustique - Graduate School (IA-GS), CNRS, Le Mans Université, France}
    
\begin{abstract}
Since the discovery of topological modes in idealized ball-and-spring kagome lattices, significant efforts have been devoted to realizing mechanical analogues of these ideal lattices via practical fabrication techniques. While numerical and experimental characterizations of these mechanical analogues have been reported, theoretical modeling that accounts for realistic structural effects (e.g., the bending behavior of thin ligaments—a departure from ideal hinges allowing free rotation) has been lacking. Here we propose a theoretical framework to investigate the dynamic properties of physically-realizable kagome metamaterials consisting of solid triangles and ligaments, in which triangles and ligaments are modeled as rigid bodies and elastic springs, respectively. By applying the framework, validated through finite element analysis, to twisted and deformed kagome metamaterials, we theoretically show the required conditions for achieving certain unique dynamic properties, including dispersion dualities and topological edge modes. The presented study unequivocally reveals the effects of structural components on these properties, which could enable new design strategies for wave propagation manipulation in kagome-based metamaterials.
\end{abstract}

	\maketitle
\section{Introduction}
Phononic crystals and mechanical metamaterials have been shown to be an excellent platform to investigate novel dynamic behavior that cannot be observed in conventional materials and structures \cite{Hussein_2014, bertoldi2017flexible,fronk2023_ND}. In addition to the discoveries of band gaps \cite{liu2000locally} and soft/localized modes \cite{tournat_njp_2011,Tournat_PRL_2011,Tournat_pre_2014,Tournat_APL_2016,Tournat_PRE_2016,Tournat_EML_2017, Tournat_EML_2017b}, a significant amount of effort has been devoted to achieving versatile wave manipulation capabilities, including acoustic diodes and switches \cite{Liang_2010,Boechler_2011,Bilal_2017,Zhou_PRB2020}, modal mixing \cite{Ganesh_2017, JIAO_JMPS_2018}, wave tuning \cite{Daraio_2006, Narisetti_2011, Jiao_prsa_2021}, negative refraction \cite{Zhu_NC_2014}, and unidirectional/nonreciprocal wave propagation \cite{Raney2016, Nadkarni_PRL_2016, Wang_PRL_2018,Moore_PRE_2018}. 

Recently, kagome-based mechanical metamaterials have attracted considerable attention due to their unique dynamic and topological properties \cite{Schaeffer2015a,paulose_2015PNAS,Ma_PRL_2018, Pishvar_PRApplied_2020, Charara_PRApplied_2021,Xiu_PNAS_2022,charara2022PNAS,Zhang_PRApplied_2023,Azizi_PRL_2023,Li_JMPS2024}, such as floppy edge modes \cite{Ma_PRL_2018,Charara_PRApplied_2021} and corner modes \cite{Zhang_PRApplied_2023,Azizi_PRL_2023}. These wave phenomena were first discovered in ideal deformed/twisted kagome lattices using ball-and-spring models, which can be classified as Maxwell lattices and are on the verge of mechanical instability, as the number of constraints is equal to the number of degrees of freedom in the unit cell \cite{Mao_Review_2018}. Sun et al. \cite{Sun_PNAS_2012} demonstrated topological boundary floppy modes in Maxwell lattices. Later, Kane and Lubensky \cite{Kane_NP_2014} introduced a polarization vector to characterize the topological floppy modes in deformed kagome lattices, relating their topological properties to the bulk.  Danawe et al. revealed the existence of corner modes localized at certain corners of finite self-dual lattices \cite{Danawe_PRB_2021}. Fruchart et al. \cite{Fruchart_Nature_2020} showed that pairs of distinct twisted kagome lattices exhibit the same band structure, which can be traced to the existence of a duality transformation between their dynamic matrices. Moreover, there exists a self-dual kagome lattice at a critical twisting angle, displaying a twofold-degenerate band structure over the entire Brillouin zone. Later, Lei et al. \cite{Lei_PRL_2022} explored this duality and  the associated hidden symmetry in 2D self-dual structures with arbitrary complexity. 

Inspired by these interesting wave phenomena in ideal kagome lattices, researchers have investigated their manifestations and potential applications in mechanical analogues of these ideal lattices. These mechanical systems can be realized via fabrication techniques, such as cutting, molding, and printing. Although the mechanical systems and the corresponding ideal lattices are highly similar in appearance (e.g., see Fig.~\ref{fig:Schematics}(a) and (d)), there exist two major physical differences. One is that triangles formed by springs and masses in ideal lattices are often replaced by triangles made of continuum solids (or thin beams) in physical specimens \cite{paulose_2015PNAS,Ma_PRL_2018,Pishvar_PRApplied_2020,charara2022PNAS,Li_JMPS2024}. The other is that these solid triangles are connected by thin ligaments instead of ideal hinges that allow free rotations in ideal lattices. Because of these differences, the theoretical models developed for ideal lattices cannot be directly applied to their mechanical counterparts, resulting in the lacking of theoretical analysis of the corresponding experimental and numerical observations that may deviate from the ``ideal"  results. For example, topological zero-frequency modes in idealized ball-and-spring lattices become finite-frequency edge modes in physical lattices \cite{Ma_PRL_2018}. Another example is that the symmetry (duality) of the phononic landscape of ideal twisted kagome lattices is broken for structural lattices of beams \cite{Gonella_PRB_2020}. It is worth noting that some theoretical treatments have been proposed to explain some of these observations, including kagome lattices with added next-nearest-neighbor
springs \cite{Stenull_PRL_2019},  ``microtwist elasticity" \cite{Nassar_JMPS_2020}, and topological elasticity in the continuum limit \cite{Sun_PRL_2020,Saremi_PRX_2020}. However, these modified treatments still rely on ideal ball-and-spring modelings.

To resolve this issue, a theoretical framework is proposed  to analytically capture the dynamic behavior of various mechanical kagome lattices. The presented framework is inspired by recent work on modeling kirigami metamaterials consisting of squares and ligaments, which can support the propagation of a variety of nonlinear waves, including elastic vector solitons \cite{Deng2017,Deng2019,ZHANG_IJSS_2023}, transition waves \cite{Korpas2021}, and their collisions \cite{Deng_nc2018,Yasuda2020,Hiromi_APL_2023,Jiao_NC_2024}. Remarkably, it has been shown that these complex wave phenomena are well captured by discrete models, in which the squares are treated as rigid bodies with translational and rotational degrees of freedom and the ligaments are modeled as equivalent elastic springs.  This modeling approach is also applied to 1D mechanical chains of triangles, theoretically revealing amplitude-dependent boundary modes \cite{Zhou_JMPS_2021} and the propagation of elastic solitons \cite{Li_IJIE_2021}. Recently, Chen et al.  \cite{CHEN_IJSS_2022} adopted this approach combined with ``microtwist elasticity"  to characterize topological
polarization in hinged kagome lattices.

Here, we focus on analytical modeling of physically-realizable kagome metamaterials, which are made of solid triangles connected by thin ligaments. The paper is organized as follows. In Section 2, a theoretical framework is introduced for kagome metamaterials with a general geometry: a discrete model is established to derive the complete equations of motion, based on which a linear wave analysis is performed to obtain dispersion relation. The analytical model is validated through finite-element analysis of regular kagome metamaterials in Section 3. The validated model is then used to explore dualities or lack thereof in twisted kagome metamaterials and topological edge modes in deformed kagome metamaterials in Section 4 and Section 5, respectively. Lastly, the work is concluded in Section 6.

\section{Theoretical framework}
\subsection{Discrete model}
\begin{figure*}[htbp]
    \centerline{ \includegraphics[width=0.85\textwidth]{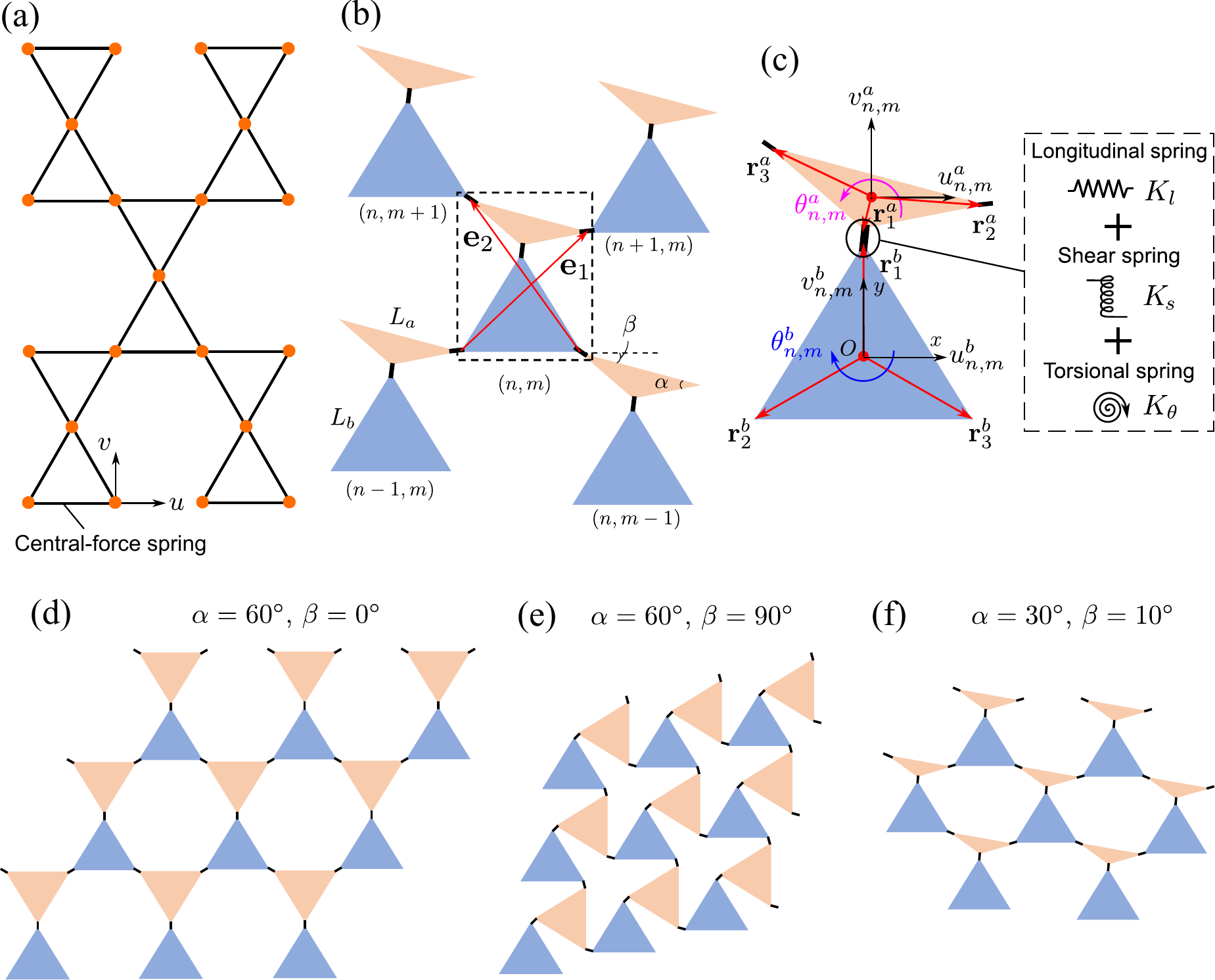}}
    \caption{Schematics of various kagome systems. (a) A ideal regular kagome lattice of point masses and central-force springs. (b) A physically-realizable kagome metamaterial of solid triangles and thin ligaments, with its unit cell highlighted by dashed black lines and theoretical modeling presented in (c): the triangles are treated as rigid bodies with two translational ($u$ and $v$) and one rotational ($\theta$) degrees of freedom and the ligaments are modeled as elastic springs through a combination of longitudinal, shear, and torsional springs. (d-f) Various kagome systems can be achieved by controlling two geometric parameters $\alpha$ and $\beta$. (d) A regular kagome metamaterial: $\alpha=60^{\circ}$, $\beta=0^{\circ}$. (e) A twisted kagome metamaterial: $\alpha=60^{\circ}$, $\beta=90^{\circ}$. (f) A deformed kagome metamaterial: $\alpha=30^{\circ}$, $\beta=10^{\circ}$.}
    \label{fig:Schematics}
\end{figure*}

To begin with, we consider a physically-realizable kagome metamaterial with a set of general geometric parameters, whose schematic is shown in Fig.~\ref{fig:Schematics}(b). $\mathbf{e}_1$ and $\mathbf{e}_2$ are the lattice vectors in real space, and $(n,m)$ is the index that identifies a unit cell in the system. The unit cell of the metamaterial, shown in Fig.~\ref{fig:Schematics}(b), is composed of two solid triangles: one equilateral and one isosceles triangles connected by a thin ligament highlighted in black. The equilateral triangle has side length $L_b$. The base length and the base angle of the isosceles triangle are denoted as $L_a$ and $\alpha$, respectively. The angle between the base and the horizontal line is denoted as $\beta$. Here, we develop a theoretical framework to investigate the dynamic behavior of such kagome metamaterial. Each isosceles (equilateral) triangle is assumed as a rigid body of mass $M_a$ ($M_b$) and rotational inertia $J_a$ ($J_b$), which has three degrees of freedom ($u$, $v$, and $\theta$). $u$ and $v$ are adopted to describe translational displacements, and $\theta$ is adopted to describe rotational angle. The stretching, shearing, and torsional behavior of the ligaments are captured using a combination of three linear springs: a longitudinal spring with stiffness $K_l$; a shear spring with stiffness $K_s$; a torsional spring with stiffness $K_\theta$. $\mathbf{r}^{a}_1=\mathbf{A}[-L_a/2  \,\,\, -L_a\tan\alpha/2]^T+\mathbf{A}[L_a/2  \,\,\, L_a\tan\alpha/6]^T$, $\mathbf{r}^{a}_2=\mathbf{A}[L_a/2  \,\,\, L_a\tan\alpha/6]^T$, and $\mathbf{r}^{a}_3=\mathbf{A}[-L_a \,\,\, 0]^T+\mathbf{A}[L_a/2  \,\,\, L_a\tan\alpha/6]^T$ are the vectors between the center of mass for the isosceles triangle and its vertices, where $\mathbf{A}=\begin{bmatrix} \cos \beta & -\sin \beta\\ \sin \beta & \cos \beta \end{bmatrix}$ is a rotation matrix and $[\cdot]^T$ denotes the transpose  of $[\cdot]$.  $\mathbf{r}^{b}_1=[0  \,\,\, L_b/\sqrt{3}]^T$, $\mathbf{r}^{b}_2=-[\cos(\pi/6)  \,\,\, \sin(\pi/6)]^TL_b/\sqrt{3}$, and $\mathbf{r}^{b}_3=[\cos(\pi/6)  \,\,\, -\sin(\pi/6)]^TL_b/\sqrt{3}$ are the vectors between the center of mass for the equilateral triangle and its vertices. Based on this discrete model, the Hamiltonian of the system can be written as 
\begin{align}\label{Hamiltonian}
\begin{split}
H&=\frac{1}{2} \sum_{n,m}\sum^{a,b}_{i} \Big[ M_i (\dot u^{i}_{n,m})^2+M_i (\dot v^{i}_{n,m})^2+J_i (\dot \theta^{i}_{n,m})^2\Big] +\sum_{n,m} \left(E^l_{n,m}+E^s_{n,m}+E^{\theta}_{n,m}\right),
\end{split}
\end{align}
where the superscript $i\in{a,b}$ and $a$ ($b$) represents the isosceles (equilateral) triangle. The first and second terms of the Hamiltonian are the kinetic energy and potential energy of the system, respectively. Specifically, the three components of the potential energy, corresponding to the strain energy stored in the three linear springs, can be expressed as 
\begin{align}\label{Potential_comp}
\begin{split}
E^l_{n,m}&=\frac{1}{2} K_l\Big[  \left(\Delta \mathbf{l}_{n,m}\cdot \mathbf{p}_1\right)^2 + \left(\Delta \mathbf{l}_{n+1,m}\cdot \mathbf{n}_1\right)^2+\left(\Delta \mathbf{l}_{n,m+1}\cdot \mathbf{m}_1 \right)^2\Big]\\
E^s_{n,m}&=\frac{1}{2} K_s\Big[  \left(\Delta \mathbf{l}_{n,m}\cdot \mathbf{p}_2\right)^2 + \left(\Delta \mathbf{l}_{n+1,m}\cdot \mathbf{n}_2\right)^2+\left(\Delta \mathbf{l}_{n,m+1}\cdot \mathbf{m}_2 \right)^2\Big]\\
E^{\theta}_{n,m}&=\frac{1}{2} K_j \Big[\left( \theta^{a}_{n,m}+\theta^{b}_{n,m}\right) ^2+(\theta^{a}_{n,m}+\theta^{b}_{n+1,m})^2+ (\theta^{a}_{n,m}+\theta^{b}_{n,m+1})^2\Big], \\
\end{split}
\end{align}
where $\Delta \mathbf{l}_{n,m}$ is the relative displacement within the ligament in unit cell $(n,m)$, and $\Delta \mathbf{l}_{n\pm1,m\pm1}$ is the relative displacement within the ligament connecting unit cells $(n,m)$ and $(n\pm1,m\pm1)$, whose full expressions are given in Appendix A. $\mathbf{n}_1$ and $\mathbf{m}_1$ are unit vectors with directions along the orientations of the ligaments connecting unit cells along the lattice vectors $\mathbf{e}_1$ and $\mathbf{e}_2$, respectively. $\mathbf{p}_1$ is a unit vector with direction along the orientation of the ligament connecting the two triangles in the unit cell. $\mathbf{n}_2$, $\mathbf{m}_2$, and $\mathbf{p}_2$ are unit vectors that are perpendicular to $\mathbf{p}_1$, $\mathbf{n}_1$, and $\mathbf{m}_1$, respectively. In the following analysis, the direction of a ligament is set to be parallel to the line passing through the centroids of two triangles that are connected by the ligament. Accordingly, these unit vectors can be expressed as 
\begin{equation}\label{Unit_vectors}
\begin{split}
\mathbf{p}_1&=\begin{bmatrix} \sin \frac{\beta}{2} \\ \cos \frac{\beta}{2}  \end{bmatrix},\,\,\,\mathbf{p}_2=\begin{bmatrix} -\cos \frac{\beta}{2} \\ \sin \frac{\beta}{2}  \end{bmatrix},\,\,\,\mathbf{n}_1=\begin{bmatrix} \cos \left(\frac{\beta}{2}-\frac{\pi}{6}\right) \\ -\sin \left(\frac{\beta}{2}-\frac{\pi}{6}\right)  \end{bmatrix},\,\,\,\mathbf{n}_2=\begin{bmatrix} \sin \left(\frac{\beta}{2}-\frac{\pi}{6}\right) \\ \cos \left(\frac{\beta}{2}-\frac{\pi}{6}\right)  \end{bmatrix},\\
\mathbf{m}_1&=\begin{bmatrix} -\cos \left(\frac{\beta}{2}+\frac{\pi}{6}\right) \\ \sin \left(\frac{\beta}{2}+\frac{\pi}{6}\right)  \end{bmatrix},\,\,\,\mathbf{m}_2=\begin{bmatrix} -\sin \left(\frac{\beta}{2}+\frac{\pi}{6}\right) \\ -\cos \left(\frac{\beta}{2}+\frac{\pi}{6}\right)  \end{bmatrix}.
\end{split}
\end{equation}

Based on the system's Hamiltonian, the equations of motion (EOMs) of the triangles in the unit cell at site $(n,m)$ can be obtained using the Hamilton’s equations: 
 \begin{align}\label{Hamiton_eqns}
\begin{split}
M_i\ddot{u}^i_{n,m}=-\frac{\partial H}{\partial u^i_{n,m}},\,\,
M_i\ddot{v}^i_{n,m}=-\frac{\partial H}{\partial v^i_{n,m}},\,\,
J_i\ddot{\theta}^i_{n,m}=-\frac{\partial H}{\partial \theta^i_{n,m}}, \,\,i\in{[a,b]}.
\end{split}
\end{align}
Substituting Eqs.~\ref{Hamiltonian} and \ref{Potential_comp} in Eq.~\ref{Hamiton_eqns} yields the EOMs of the triangles in the unit cell at site $(n,m)$. For the isosceles triangle, the EOMs have the form 
\begin{equation}
\begin{split}
	\mathbf{M}_a\ddot{\mathbf{u}}^a_{n,m}
	&=-  (K_l \mathbf{p}_1\otimes\mathbf{p}_1+K_s\mathbf{p}_2\otimes\mathbf{p}_2) \Delta \mathbf{l}_{n,m} +  (K_l \mathbf{n}_1\otimes\mathbf{n}_1+K_s\mathbf{n}_2\otimes\mathbf{n}_2)\Delta \mathbf{l}_{n+1,m}\\
	&+  (K_l \mathbf{m}_1\otimes\mathbf{m}_1+K_s\mathbf{m}_2\otimes\mathbf{m}_2)\Delta \mathbf{l}_{n,m+1}, \\
\end{split}
\end{equation}
where $\mathbf{u}^{a}_{n,m}=\begin{bmatrix} u^{a}_{n,m} \\ v^{a}_{n,m} \end{bmatrix}$, $\mathbf{M}_a=\begin{bmatrix} M_a & 0\\ 0 & M_a \end{bmatrix}$ is a mass matrix,  and $\otimes$ denotes the dyadic product.
\begin{equation}
\begin{split}
J_a\ddot{\theta}^a_{n,m}&=-K_j(\theta^a_{n,m}+\theta^b_{n,m})  -K_j(2\theta^a_{n,m}+\theta^b_{n+1,m}+\theta^b_{n,m+1})-  \left( K_l\mathbf{p}_1^T\mathbf{C'}_{n,m}^{a}\mathbf{r}_{1}^a\mathbf{p}_1+K_s\mathbf{p}_2^T\mathbf{C'}_{n,m}^{a}\mathbf{r}_{1}^a\mathbf{p}_2\right) \cdot\Delta \mathbf{l}_{n,m}\\
&+ \left( K_l\mathbf{n}_1^T\mathbf{C'}_{n,m}^{a}\mathbf{r}_{2}^a\mathbf{n}_1+K_s\mathbf{n}_2^T\mathbf{C'}_{n,m}^{a}\mathbf{r}_{2}^a\mathbf{n}_2\right) \cdot\Delta \mathbf{l}_{n+1,m} +  \left( K_l\mathbf{m}_1^T\mathbf{C'}_{n,m}^{a}\mathbf{r}_{3}^a\mathbf{m}_1+K_s\mathbf{m}_2^T\mathbf{C'}_{n,m}^{a}\mathbf{r}_{3}^a\mathbf{m}_2\right) \cdot\Delta \mathbf{l}_{n,m+1},\\
\end{split}
\end{equation}
where $\mathbf{C'}_{n,m}^{a}=\begin{bmatrix} -\sin \theta^a_{n,m} & -\cos \theta^a_{n,m}\\ \cos \theta^a_{n,m} & -\sin \theta^a_{n,m} \end{bmatrix}$.

For the equilateral triangle, the EOMs are given as
\begin{equation}
\begin{split}
	\mathbf{M}_b\ddot{\mathbf{u}}^b_{n,m}
	&=(K_l \mathbf{p}_1\otimes\mathbf{p}_1+K_s\mathbf{p}_2\otimes\mathbf{p}_2)\Delta \mathbf{l}_{n,m} +  (K_l \mathbf{n}_1\otimes\mathbf{n}_1+K_s\mathbf{n}_2\otimes\mathbf{n}_2)\Delta \mathbf{l}_{n-1,m}  \\
	&+  (K_l \mathbf{m}_1\otimes\mathbf{m}_1+K_s\mathbf{m}_2\otimes\mathbf{m}_2)\Delta \mathbf{l}_{n,m-1}, \\
\end{split}
\end{equation}
\begin{equation}
\begin{split}
J_b\ddot{\theta}^b_{n,m}&=-K_j (\theta^a_{n,m}+\theta^b_{n,m})  -K_j(2\theta^b_{n,m}+\theta^a_{n-1,m}+\theta^a_{n,m-1})+\left( K_l\mathbf{p}_1^T\mathbf{C'}_{n,m}^{b}\mathbf{r}_{1}^b\mathbf{p}_1+K_s\mathbf{p}_2^T\mathbf{C'}_{n,m}^{b}\mathbf{r}_{1}^b\mathbf{p}_2\right) \cdot\Delta \mathbf{l}_{n,m}\\
&+ \left( K_l\mathbf{n}_1^T\mathbf{C'}_{n,m}^{b}\mathbf{r}_{2}^b\mathbf{n}_1+K_s\mathbf{n}_2^T\mathbf{C'}_{n,m}^{b}\mathbf{r}_{2}^b\mathbf{n}_2\right) \cdot\Delta \mathbf{l}_{n-1,m}  +  \left( K_l\mathbf{m}_1^T\mathbf{C'}_{n,m}^{b}\mathbf{r}_{3}^b\mathbf{m}_1+K_s\mathbf{m}_2^T\mathbf{C'}_{n,m}^{b}\mathbf{r}_{3}^b\mathbf{m}_2\right) \cdot\Delta \mathbf{l}_{n,m-1},\\
\end{split}
\end{equation}
where $\mathbf{u}^{b}_{n,m}=\begin{bmatrix} u^{b}_{n,m} \\ v^{b}_{n,m} \end{bmatrix}$, $\mathbf{M}_b=\begin{bmatrix} M_b & 0\\ 0 & M_b \end{bmatrix}$ and $\mathbf{C'}_{n,m}^{b}=\begin{bmatrix} -\sin \theta^b_{n,m} & \cos \theta^b_{n,m}\\ -\cos \theta^b_{n,m} & -\sin \theta^b_{n,m} \end{bmatrix}$.

\subsection{Linear wave analysis}

To investigate the propagation of linear waves in such kagome metamaterials, we first need to linearize the above EOMs. Under the assumption of $\theta\ll 1$ for linear waves, we employ the approximation that $\cos \theta \approx 1$ and $\sin \theta \approx\theta$ to derive the linearized EOMs, which can be written in the following matrix form:
\begin{equation}\label{Govering_eqn}
\mathbf{M}\ddot{\mathbf{U}}_{n,m}+\mathbf{K}_1\mathbf{U}_{n,m}+\mathbf{K}_2\mathbf{U}_{n-1,m}+\mathbf{K}_3\mathbf{U}_{n+1,m}+\mathbf{K}_4\mathbf{U}_{n,m-1}+\mathbf{K}_5\mathbf{U}_{n,m+1}=\mathbf{0},
\end{equation}
where $\mathbf{U}_{n\pm1,m\pm1}=[u^{a}_{n\pm1,m\pm1}\,\, u^{a}_{n\pm1,m\pm1} \,\,\theta^{a}_{n\pm1,m\pm1}\,\, u^{b}_{n\pm1,m\pm1}\,\, u^{b}_{n\pm1,m\pm1} \,\,\theta^{b}_{n\pm1,m\pm1}]^T$ and the full expressions of $\mathbf{M}$ and $\mathbf{K}_i,\,\, i\in[1,2,3,4,5]$ are given in Appendix A. Under plane wave assumptions, $\mathbf{U}_{n,m}$ can be written in the form
\begin{equation}\label{Govering_sln}
\mathbf{U}_{n,m}=A \mathbf{\Phi} e^{i(\mathbf{k} \cdot \mathbf{r}_{n,m}-\omega t)},
\end{equation}
where $A$ is an amplitude constant, $\mathbf{\Phi}$ is a modal vector, $\mathbf{r}_{n,m}$ is the position vector of the unit cell at site $(n,m)$, and $\mathbf{k}$ and $\omega$ are the wave vector and angular frequency of a traveling wave, respectively. Based on Floquet-Bloch theorem, the relation between displacements at neighboring sites are given as
\begin{equation}\label{Govering_Bloch}
\mathbf{U}_{n\pm1,m\pm1}=\mathbf{U}_{n,m}e^{i(\pm\mathbf{k} \cdot \mathbf{e}_1\pm\mathbf{k} \cdot \mathbf{e}_2)}.
\end{equation}
Substituting Eq.~\ref{Govering_sln} and Eq.~\ref{Govering_Bloch} in Eq.~\ref{Govering_eqn} yields the following eigenvalue problem
\begin{equation}\label{Govering_eigen}
\Big[-\omega^2\mathbf{M}+\mathbf{K}(\mathbf{k})\Big]\mathbf{\Phi}=\mathbf{0},
\end{equation}
where 
\begin{equation}\label{K_matrix}
\begin{split}
\mathbf{K}(\mathbf{k})=\mathbf{K}_1e^{-i\mathbf{k} \cdot \mathbf{e}_1}+\mathbf{K}_2e^{i\mathbf{k} \cdot \mathbf{e}_1}+\mathbf{K}_3e^{-i\mathbf{k} \cdot \mathbf{e}_2}+\mathbf{K}_4e^{i\mathbf{k} \cdot \mathbf{e}_2}+\mathbf{K}_5\equiv\begin{bmatrix} k_{11} & k_{12} & k_{13}\\ k_{21} & k_{22} & k_{23}\\k_{31} & k_{32} & k_{33} \end{bmatrix}
\end{split}
\end{equation}
is a wave-vector-dependent stiffness matrix and its components $k_{ij}$ are $2\times2$ matrices. For a set of parameters, we can obtain the dispersion relation by solving Eq.~\ref{Govering_eigen}. In the next section, we will rely on the presented theoretical model to investigate the propagation of linear waves in regular, twisted, and topological kagome metamterials. 

\section{Finite element validation}

\begin{figure*}[htbp]
    \centerline{ \includegraphics[width=0.9\textwidth]{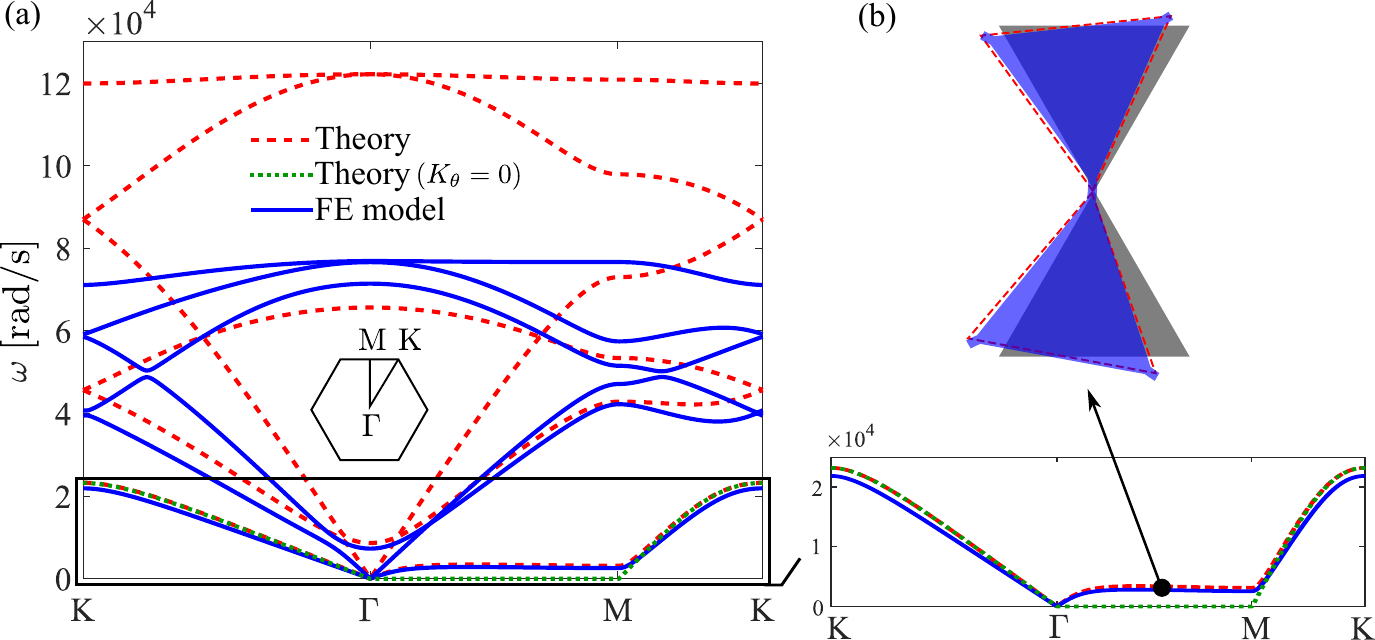}}
    \caption{(a) Band diagrams of a regular kagome metamaterial obtained from the presented theoretical model and finite element (FE) analysis. The red dashed curves are theoretically predicted using  parameters determined from FE simulations, while the dotted green curve is obtained by artificially setting $K_\theta=0$. The blue curves are obtained from FE analysis. A zoomed view of the first bands is shown in the inset. (b) Mode shapes calculated at the highlighted point along the $\Gamma-\mathrm{M} $ direction: theoretical prediction (dashed red lines) vs FE result (blue domain). The grey domain presents the undeformed unit cell.} 
    \label{fig:Banddiagram_regular}
\end{figure*}

By setting $L_a=L_b$, $\alpha=60^{\circ}$, and $\beta=0^{\circ}$, the system reduces to a regular kagome metamaterial, as shown in Fig.~\ref{fig:Schematics}(d). To validate the discrete model, we perform finite element (FE) analysis on a continuum kagome system with thickness $0.1$ cm, whose unit cell consists of two identical equilateral triangles with side length $L_a=L_b=2$ cm. The ligament between the two triangles has the dimension of width $0.1$~cm $\times$ length 0.25 cm. The 2D body is discretized using quadrilateral plane stress elements, and Bloch conditions are applied along lattice vectors (see details of the FE model in Appendix B). The material adopted in the FE model is acrylonitrile butadiene styrene (ABS) with the following material parameters: Young's modulus $E=2.14$ Gpa, Poisson's ratio $\nu=0.35$, and density $\rho=1040$ Kg/m$^3$. 

In Fig.~\ref{fig:Banddiagram_regular}(a), we plot the dispersion relation (blue curves) calculated from the FE model. To validate our analytical model, we predict the dispersion relation by solving the corresponding eigenvalue problem (i.e., Eq.~\ref{Govering_eigen}). Using ABS material parameters, we determine $M_a=M_b\approx1.8\times10^{-4}$ kg and $J_a=J_b=M_aL_a^2/12\approx6\times10^{-9}$ kg$\cdot$m$^2$. The three equivalent spring stiffnesses of the ligament are determined as $K_l\approx8.64\times10^5$ N/m, $K_s\approx3.24\times10^4$ N/m, and $K_\theta\approx7.45\times10^{-2}$ N$\cdot$m/rad via simulating the static responses of one ligament under three sets of boundary conditions (see Fig.S1(b) in Appendix B for details). With these parameters, we obtain the analytically-predicted dispersion relation, which is displayed in Fig.~\ref{fig:Banddiagram_regular}(a) as dashed red curves.

Comparing the theoretical result with the FE model, we observe an excellent agreement between the two first branches, especially in the $\Gamma$-M direction. However, the two band diagrams differ more substantially as frequency increases, which can be ascribed to the fact that, in contrast to the rigid body assumptions in the theoretical framework, the FE model allows deformations occurring within the triangles in higher-frequency modes (and thereby softening the whole band diagram). In Fig.~\ref{fig:Banddiagram_regular}(b), we show the mode shape predicted from the theory (red dashed lines) at the highlighted point in the plateau region of the first branch, matching well with the mode shape obtained from the FE model (blue area). 

Now, we rely on the analytical model to investigate the effects of thin ligaments on the first branch. If $K_\theta$ is artificially set to zero, the first branch along the $\Gamma$-M direction drops to zero frequency, as indicated by the green dotted curve in Fig.~\ref{fig:Banddiagram_regular}(a), analogous to the isostatic
zero-frequency branch in ideal regular kagome lattices. This implies that the torsional/bending stiffness induced by the ligaments results in a shift-up toward finite frequencies of the zero-frequency branch. Finally, we note that the presented findings are qualitatively consistent with previous experimental results on similar kagome metamaterials \cite{Ma_PRL_2018}.  In the following sections, we employ the validated discrete model to investigate the effects of different parameters on dispersion dualities in twisted kagome and edge modes in topological kagome lattices. In the following analyses, the values of the parameters are arbitrarily chosen, and standard SI units are adopted and omitted  for simplicity.

\section{Dispersion Dualities in Twisted kagome Metamaterials}
A twisted kagome matermaterial can be obtained by setting $L_a=L_b$, $\alpha=60^{\circ}$, and choosing a twisting angle $\beta\neq0^{\circ}$. For ideal twisted kagome lattices, it has been shown that pairs of distinct configurations exhibit identical dispersion relation, which is linked to the existence of a mathematical duality between the stiffness matrices of pairs of kagome lattices. Specifically, for a twisted kagome lattice with angle $\bar{\beta}$, there exists a dual kagome lattice with angle $\beta^*=2\beta_c-\bar{\beta}$, where $\beta_c=\pi/2$ is the critical point. Here, we employ the proposed theoretical framework to explore whether such dispersion duality can be preserved in our twisted kagome metamaterials.

\begin{figure*}[htbp]
    \centerline{ \includegraphics[width=0.85\textwidth]{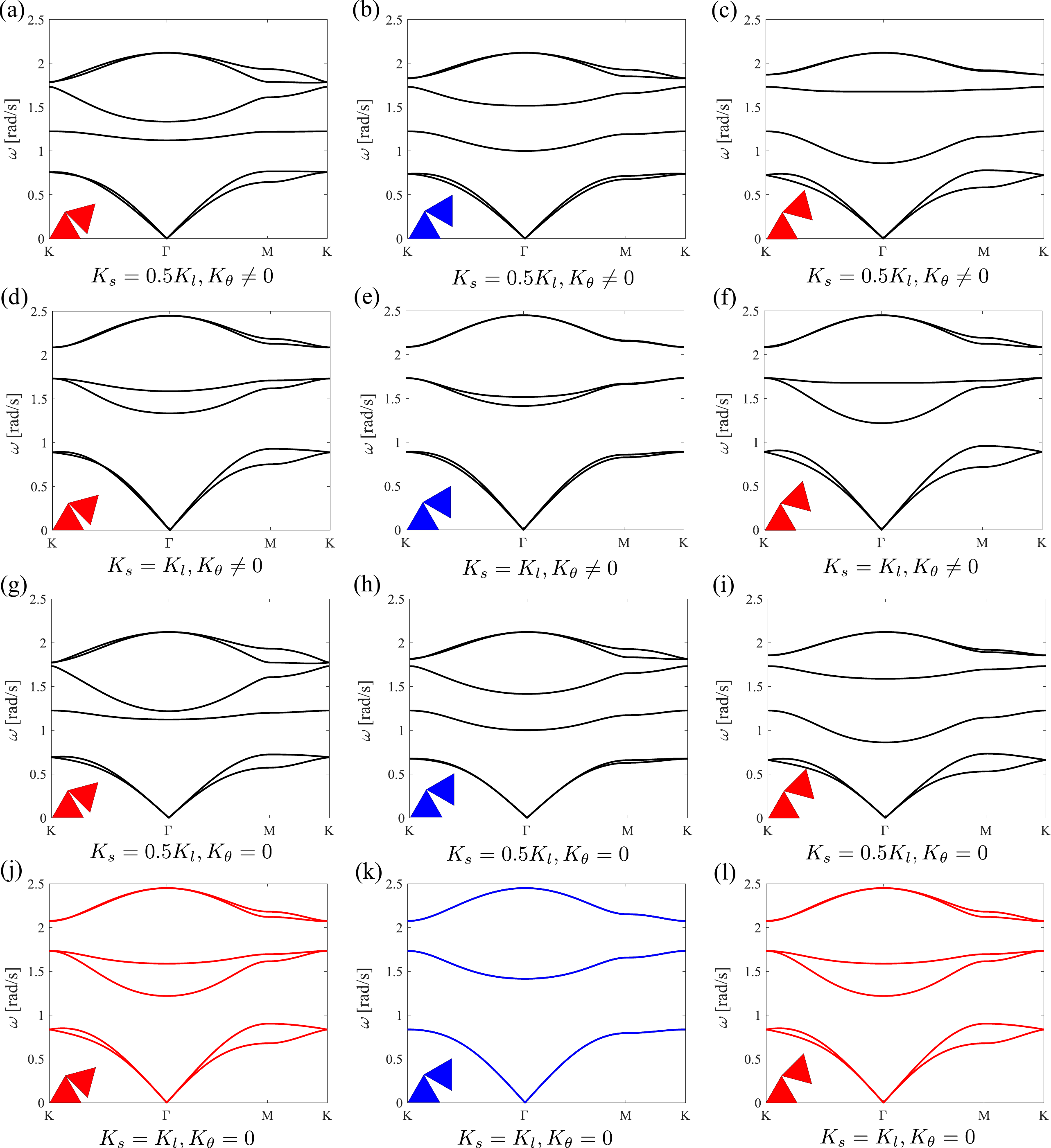}}
    \caption{Band diagrams for a self-dual kagome metamaterial with $\beta_c=90^{\circ}$ and a pair of twisted kagome metamaterials with $(\bar{\beta}, \beta^*)=(75^{\circ}, 105^{\circ})$ for four different choices of spring stiffnesses: (a-c) $K_s=0.5K_l$ and $K_\theta\neq0$; (d-f) $K_s=K_l$ and $K_\theta\neq0$; (g-i) $K_s=0.5K_l$ and $K_\theta=0$; (j-l) $K_s=K_l$ and $K_\theta=0$. Only in the case for $K_s=K_l$ and $K_\theta=0$, the dispersion duality between the paired systems is preserved, and a twofold-degeneracy over the entire Brillouin contour is observed in the self-dual kagome metamaterial.} 
    \label{fig:Band_diagram_Twisted}
\end{figure*}

We consider a pair of twisted kagome metamaterials with angles $(\bar{\beta}, \beta^*)=(75^{\circ}, 105^{\circ})$ and the following parameters: $M_a=M_b=1$, $J_a=J_b=2\times10^{-4}$, $L_a=L_b=2\times10^{-2}$, and $K_l=1$, and $K_\theta=1\times10^{-5}$ (if not zero). To study the effects of the ligaments, we consider four different choices of $K_s$ and $K_\theta$: (i) $K_s=0.5K_l$ and $K_\theta\neq0$; (ii) $K_s=K_l$ and $K_\theta\neq0$; (iii) $K_s=0.5K_l$ and $K_\theta=0$; (iv) $K_s=K_l$ and $K_\theta=0$. In Fig.~\ref{fig:Band_diagram_Twisted}, we report the dispersion relations for the pair of twisted kagome metamaterials using the four sets of parameters. In addition, we report the results for a kagome metamaterial with the critical twisting angle $\beta_c=90^{\circ}$, as shown in Fig.~\ref{fig:Schematics}(e). Interestingly, we observe in Fig.~\ref{fig:Band_diagram_Twisted}(j) and (l) that the two systems with angles $\bar{\beta}$ and $\beta^*$ have the same dispersion relation only if $K_s=K_l$ and $K_\theta=0$. For all other three cases, the three systems exhibit noticeably distinct dispersive characteristics. Moreover, for the self-dual twisted kagome system with angle $\beta_c$, the dispersion relation is doubly degenerate over the
entire irreducible Brillouin zone only if $K_s=K_l$ and $K_\theta=0$, as shown in Fig.~\ref{fig:Band_diagram_Twisted}(k). Remarkably, we note that these findings are highly consistent with the numerical results reported previously \cite{Gonella_PRB_2020}.

Based on the above observations, we now prove that the eigenvalue problems for any pair of twisted kagome metamaterials with angle $\bar{\beta}$ and $\beta^*=2\beta_c-\bar{\beta}$ yield identical $\omega-\mathbf{k}$ results when $K_s=K_l$ and $K_\theta=0$. Since the mass matrix $\mathbf{M}$ is a diagonal matrix consisting of constants and the pair of stiffness matrices $\mathbf{K}(\bar{\beta},\mathbf{k})$ and $\mathbf{K}(\beta^*,\mathbf{k})$ have the same diagonal components (i.e., $K_{ii}(\bar{\beta},\mathbf{k})=K_{ii}(\beta^*,\mathbf{k}), i\in[1, 2,3,4,5,6]$), the problem reduces to whether we can prove $\det(\mathbf{K}(\bar{\beta},\mathbf{k}))=\det(\mathbf{K}(\beta^*,\mathbf{k}))$. For a stiffness matrix $\mathbf{K}(\beta,\mathbf{k})$ whose form is given by Eq~\ref{K_matrix}, its determinant can be calculated as 
\begin{equation}\label{K_determinant}
\det(\mathbf{K})=\det\left(k_{11}-\begin{bmatrix} k_{12} & k_{13}\end{bmatrix}\begin{bmatrix} k_{22} & k_{23}\\ k_{32} & k_{33}\end{bmatrix}^{-1}\begin{bmatrix} k_{21} \\ k_{31}\end{bmatrix}\right)\det\left(k_{22}-k_{23}k_{33}^{-1}k_{32}\right)\det(k_{33}).
\end{equation}
Substituting $K_s=K_l$, $K_\theta=0$, and Eq.~\ref{Unit_vectors} in the analytical expressions of $k_{ij}$ yields 
\begin{equation}\label{K_comps_relation}
\begin{split}
k_{11}&=k_{22}=K_l\begin{bmatrix} 3 & 0\\ 0 & 3\end{bmatrix}, \,\,\,k_{12}=-K_l\begin{bmatrix} 1+e^{i\mathbf{k}\cdot\mathbf{e}_1}+e^{i\mathbf{k}\cdot\mathbf{e}_2} & 0\\ 0 & 1+e^{i\mathbf{k}\cdot\mathbf{e}_1}+e^{i\mathbf{k}\cdot\mathbf{e}_2}\end{bmatrix}, \\
k_{13}&=\frac{K_lL_a}{\sqrt{3}}\begin{bmatrix} 0 & -1+\sin{\frac{\pi}{6}}\left(e^{i\mathbf{k}\cdot\mathbf{e}_1}+e^{i\mathbf{k}\cdot\mathbf{e}_2}\right)\\ 0 & \cos{\frac{\pi}{6}}\left(-e^{i\mathbf{k}\cdot\mathbf{e}_1}+e^{i\mathbf{k}\cdot\mathbf{e}_2}\right)\end{bmatrix},\\
k_{21}&=-K_l\begin{bmatrix} 1+e^{-i\mathbf{k}\cdot\mathbf{e}_1}+e^{-i\mathbf{k}\cdot\mathbf{e}_2} & 0\\ 0 & 1+e^{-i\mathbf{k}\cdot\mathbf{e}_1}+e^{-i\mathbf{k}\cdot\mathbf{e}_2}\end{bmatrix},\\k_{23}&=-\frac{K_lL_a}{\sqrt{3}}\begin{bmatrix} \cos{\beta}+\sin(\beta-\frac{\pi}{6})e^{-i\mathbf{k}\cdot\mathbf{e}_1}-\sin(\beta+\frac{\pi}{6})e^{-i\mathbf{k}\cdot\mathbf{e}_2} & 0\\ -\sin{\beta}+\cos(\beta-\frac{\pi}{6})e^{-i\mathbf{k}\cdot\mathbf{e}_1}-\cos(\beta+\frac{\pi}{6})e^{-i\mathbf{k}\cdot\mathbf{e}_2} & 0\end{bmatrix},\\k_{31}&=\frac{K_lL_a}{\sqrt{3}}\begin{bmatrix} 0 & 0\\ -1+\sin{\frac{\pi}{6}}\left(e^{-i\mathbf{k}\cdot\mathbf{e}_1}+e^{-i\mathbf{k}\cdot\mathbf{e}_2}\right) & \cos{\frac{\pi}{6}}\left(-e^{-i\mathbf{k}\cdot\mathbf{e}_1}+e^{-i\mathbf{k}\cdot\mathbf{e}_2}\right)\end{bmatrix},\\
k_{32}&=-\frac{K_lL_a}{\sqrt{3}}\begin{bmatrix} \cos{\beta}+\sin(\beta-\frac{\pi}{6})e^{i\mathbf{k}\cdot\mathbf{e}_1}-\sin(\beta+\frac{\pi}{6})e^{i\mathbf{k}\cdot\mathbf{e}_2} & -\sin{\beta}+\cos(\beta-\frac{\pi}{6})e^{i\mathbf{k}\cdot\mathbf{e}_1}-\cos(\beta+\frac{\pi}{6})e^{i\mathbf{k}\cdot\mathbf{e}_2}\\ 0 & 0\end{bmatrix},\\
k_{33}&=\frac{K_lL_a^2}{3}\begin{bmatrix} 3 & -\cos{\beta}\left(1+e^{i\mathbf{k}\cdot\mathbf{e}_1}+e^{i\mathbf{k}\cdot\mathbf{e}_2}\right)\\ -\cos{\beta}\left(1+e^{-i\mathbf{k}\cdot\mathbf{e}_1}+e^{-i\mathbf{k}\cdot\mathbf{e}_2}\right) & 3\end{bmatrix}.
\end{split}
\end{equation}
From Eq.~\ref{K_determinant} and Eq.~\ref{K_comps_relation}, one can verify that, for a dual pair of twisting angles $\bar{\beta}$ and $\beta^*=2\beta_c-\bar{\beta}$, $\det(\mathbf{K}(\bar{\beta},\mathbf{k}))=\det(\mathbf{K}(\beta^*,\mathbf{k}))$. Thus, we have shown that a twisted kagome metamaterial and its dual system have the same dispersion relation. This is mathematically equivalent to $\mathbf{K(\bar{\beta},\mathbf{k})}=\mathbf{M}\mathbf{P}\mathbf{V}\mathbf{P}^{-1}$ and $\mathbf{K(\beta^*,\mathbf{k})}=\mathbf{M}\mathbf{Q}\mathbf{V}\mathbf{Q}^{-1}$, where $\mathbf{V}$ is a diagonal matrix of eigenvalues $\omega$, and $\mathbf{P}$ and $\mathbf{Q}$ consist of the corresponding eigenvectors $\mathbf{\Phi}$. By introducing $\Tilde{\mathbf{K}}=\mathbf{M}^{-1}\mathbf{K}$ and $\mathbf{D}=\mathbf{P}\mathbf{Q}^{-1}$, we can easily obtain the duality between $\Tilde{\mathbf{K}}(\bar{\beta},\mathbf{k})$ and $\Tilde{\mathbf{K}}(\beta^*,\mathbf{k})$
\begin{equation}\label{K_duality}
\Tilde{\mathbf{K}}(\bar{\beta},\mathbf{k})=\mathbf{D}\Tilde{\mathbf{K}}(\beta^*,\mathbf{k})\mathbf{D}^{-1},
\end{equation}
Interestingly, we observe that Eq.~\ref{K_duality} has the form consistent with that derived for ideal twisted kagome lattices~\cite{Fruchart_Nature_2020}. However, the duality in our systems is achieved only for  case (iv), i.e., $K_s=K_l$ and $K_\theta=0$. In practice, these conditions can be possibly established by carefully designing the interactions between solid triangles. A promising strategy is to leverage magnetic interactions, as they can provide uniaxial forces and possibly avoid inducing any torsional force. Moreover, a direct calculation shows that the stiffness matrix $\mathbf{K}(\beta_c,\mathbf{k})$ at the critical point $\beta_c=\pi/2$ leads to the global twofold degeneracy observed in Fig.~\ref{fig:Band_diagram_Twisted}(k). Based on these findings, we have shown that the theoretical model can be used to investigate the effects of different parameters on the existence of dualities, thus providing useful guidelines for the design of physically-realizable twisted kagome metamaterials.

\section{Topological edge modes in deformed kagome Metamaterials}

\begin{figure*}[htbp]
    \centerline{ \includegraphics[width=1\textwidth]{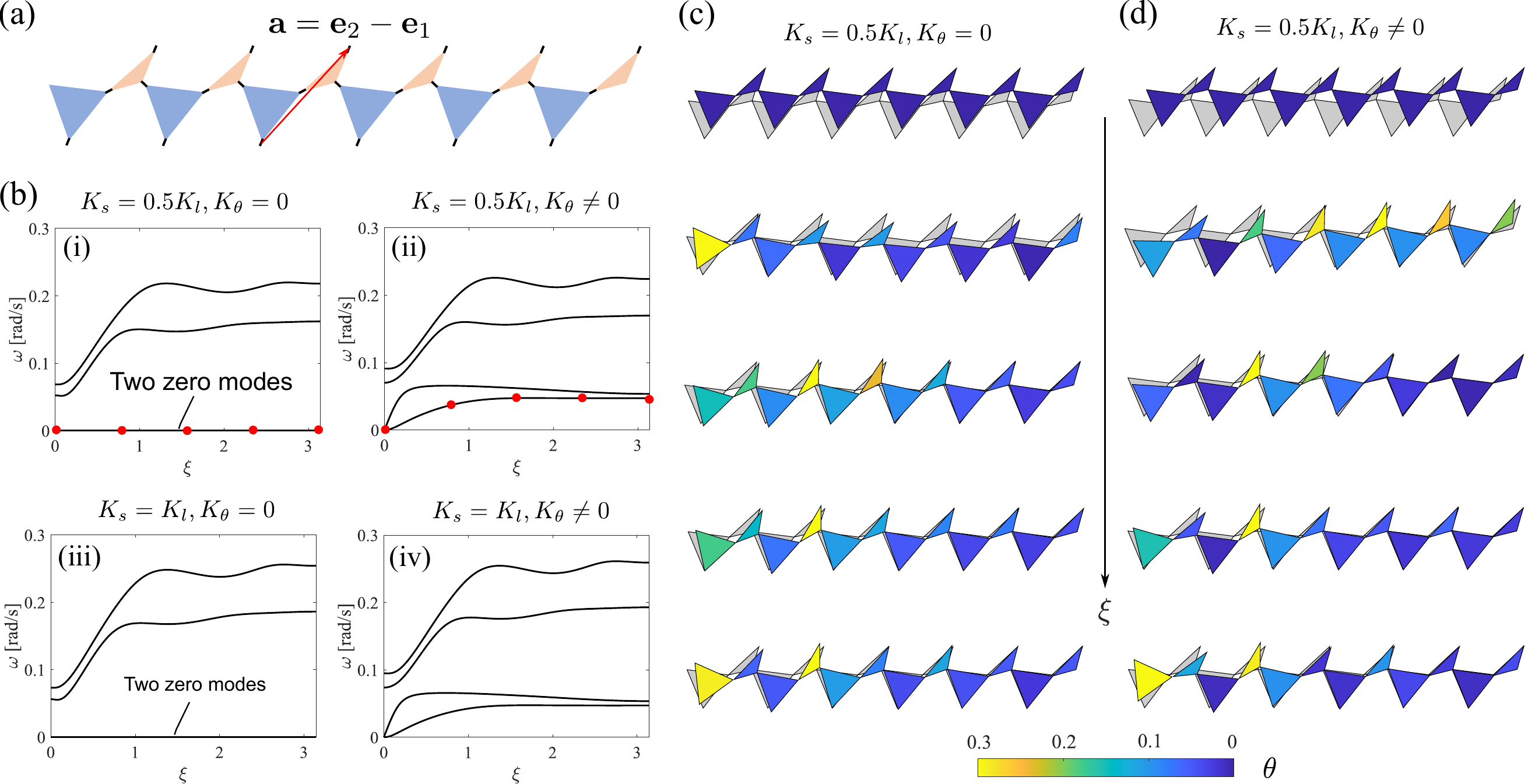}}
    \caption{Topological edge modes in deformed kagome metamaterials. (a) Schematic of a supercell of six unit cells. Bloch conditions are applied along vector $\mathbf{a}$, and the left and right edges are free. (b) Band diagrams of the supercell for four different choices of spring stiffnesses: (i) $K_s=0.5K_l$ and $K_\theta=0$; (ii) $K_s=0.5K_l$ and $K_\theta\neq0$; (iii) $K_s=K_l$ and $K_\theta=0$; (iv) $K_s=K_l$ and $K_\theta\neq0$. Two zero modes are revealed in cases (i) and (iii), while two edge modes at finite frequencies are revealed in cases (ii) and (iv). (c-d) Mode shapes calculated along (c) one zero mode found in case (i) and along the lowest edge mode found in case (ii) at selected wavenumbers indicated by red dots in (b).}
    \label{fig:Supercell_analysis}
\end{figure*}

In this section, we aim to investigate how topological zero modes manifest in the presented kagome systems (translations of the whole structure are trivial zero modes). For an arbitrary ideal lattice with $N$ mass points and $N_b$ bonds (i.e., central-force springs), there may exist $N_0$ zero modes and $N_{s}$ self-stress modes. According to the index theorem \cite{Kane_NP_2014}, $N_0$ and $N_{s}$ obeys the relation
\begin{equation}\label{Maxwell_relation}
N_0-N_{s}=dN-N_b,
\end{equation}
where $d$ denotes the dimension of the lattice (e.g., $d=2$ for 2D ideal systems). A special class of lattices is referred to as Maxwell lattices in which the average coordination number $z$ exactly equals to $2d$ under periodic boundary conditions. Based on this definition, ideal square and kagome lattices can be classified as 2D Maxwell lattices. Moreover, it follows from Eq.~\ref{Maxwell_relation} and $N_b=zN/2$ that $N_0=N_{s}$ is true for Maxwell lattices. Zero modes (e.g., Guest-Hutchinson modes \cite{GUEST_JMPS_2003,LI_JMPS_2023})  can be determined by finding the null space of the compatibility matrix $\mathbf{C}$, which relates the displacements of the masses $\mathbf{u}$ to the extensions in the bonds $\Delta \mathbf{e}$, i.e., $\Delta \mathbf{l}=\mathbf{C}\mathbf{u}$. To describe the topological properties of these zero modes, Kane and Lubensky \cite{Kane_NP_2014} introduced a topological polarization vector $\mathbf{R}_T=-\sum_{i=1}n_i\mathbf{e}_i$, where $n_i$ are winding numbers calculated from 
\begin{equation}\label{Topological polarization}
n_i=\frac{1}{2\pi}\oint_{C_i}d\mathbf{k}\cdot \Delta_{\mathbf{k}}\ln[\det\mathbf{C(\mathbf{k})}].
\end{equation}
 Here $\mathbf{C(\mathbf{k})}$ is the compatibility matrix in Fourier space, and $C_i$ is a cycle of the Brillouin zone connecting $\mathbf{k}$ and $\mathbf{k}+\mathbf{q}_i$, where $\mathbf{q}_i$ is a primitive reciprocal vector satisfying $\mathbf{q}_i\cdot \mathbf{e}_j=2\pi\delta_{ij}$. 

Since the above topological characterization is developed for Maxwell lattices, we first need to determine whether our kagome systems belong to Maxwell lattices. Despite the appearance similarities between our systems and ideal kagome lattices, they have distinct mathematical modeling as demonstrated in Sec. II. Specifically, each triangle has three degrees of freedom ($u$, $v$ and $\theta$), and each bond consists of three springs (longitudinal, shear and torsional). In the unit cell, there are two triangles and three bonds, so the number of degrees of freedom $n_d$ and the number of constraints $n_c$ per unit cell are $n_d=6$ and $n_c=9\neq n_d$, respectively. Interestingly, if we set $K_{\theta}=0$, the three torsional constraints in the unit cell are relaxed, leading to $n_d=n_c=6$ (note that the torsional degrees of freedom do not vanish). Based on the definition, the system now becomes a Maxwell lattice, which we refer to as the reference system.

Recently, Ma et al. \cite{Ma_PRL_2018} experimentally revealed that the zero-energy floppy edge modes of ideal topological lattices morph into edge modes at finite frequencies in physical specimens. This phenomenon can be empirically ascribed to the existence of ligaments that provide
bending/torsional stiffness between triangles in the specimens. Here, we employ the proposed  theoretical model to better understand this experimental observation. To this end, we choose a set of geometric parameters ($L_a=L_b=1$, $\alpha=30^{\circ}$, $\beta=10^{\circ}$) that yield a lattice geometry, as shown in Fig.~\ref{fig:Schematics}(f), consistent with the one studied in \cite{Ma_PRL_2018}. We first investigate the topological properties of the corresponding reference system by setting $K_{\theta}=0$. The compatibility matrix can be expressed as
\begin{equation}\label{C_matrix}
\begin{split}
\mathbf{C}(\mathbf{k})=\begin{bmatrix} \mathbf{p}_1(1) & \mathbf{p}_1(2) & -\mathbf{p}_1(1) & -\mathbf{p}_1(2) & \mathbf{p}_1\cdot \mathbf{g}_1^a & \mathbf{p}_1\cdot \mathbf{g}_1^b\\ \mathbf{p}_2(1) & \mathbf{p}_2(2) & -\mathbf{p}_2(1) & -\mathbf{p}_2(2) & \mathbf{p}_2\cdot \mathbf{g}_1^a & \mathbf{p}_2\cdot \mathbf{g}_1^b\\-\mathbf{n}_1(1) & -\mathbf{n}_1(2) & \mathbf{n}_1(1)e^{i\mathbf{k} \cdot \mathbf{e}_1} & \mathbf{n}_1(2)e^{i\mathbf{k} \cdot \mathbf{e}_1} & -\mathbf{n}_1\cdot \mathbf{g}_2^a & -\mathbf{n}_1\cdot \mathbf{g}_2^be^{i\mathbf{k} \cdot \mathbf{e}_1}\\-\mathbf{n}_2(1) & -\mathbf{n}_2(2) & \mathbf{n}_2(1)e^{i\mathbf{k} \cdot \mathbf{e}_1} & \mathbf{n}_2(2)e^{i\mathbf{k} \cdot \mathbf{e}_1} & -\mathbf{n}_2\cdot \mathbf{g}_2^a & -\mathbf{n}_2\cdot \mathbf{g}_2^be^{i\mathbf{k} \cdot \mathbf{e}_1}\\-\mathbf{m}_1(1) & -\mathbf{m}_1(2) & \mathbf{m}_1(1)e^{i\mathbf{k} \cdot \mathbf{e}_2} & \mathbf{m}_1(2)e^{i\mathbf{k} \cdot \mathbf{e}_2} & -\mathbf{m}_1\cdot \mathbf{g}_3^a & -\mathbf{m}_1\cdot \mathbf{g}_3^be^{i\mathbf{k} \cdot \mathbf{e}_2}\\-\mathbf{m}_2(1) & -\mathbf{m}_2(2) & \mathbf{m}_2(1)e^{i\mathbf{k} \cdot \mathbf{e}_2} & \mathbf{m}_2(2)e^{i\mathbf{k} \cdot \mathbf{e}_2} & -\mathbf{m}_2\cdot \mathbf{g}_3^a & -\mathbf{m}_2\cdot \mathbf{g}_3^be^{i\mathbf{k} \cdot \mathbf{e}_2}\end{bmatrix},
\end{split}
\end{equation}
where $\mathbf{g}_i^{a}=[-\mathbf{r}^a_i(2) \,\,\, \mathbf{r}^a_i(1)]^T$ and $\mathbf{g}_i^{b}=[-\mathbf{r}^{b}_i(2) \,\,\, \mathbf{r}^{b}_i(1)]^T, \,\,\, i\in[1,2,3]$. From Eq.~\ref{Topological polarization}, we get the topological polarization vector $\mathbf{R}_T=\mathbf{e}_2$ with direction pointing to the top edge. This indicates the existence of topological zero-energy floppy modes occurring at the top edge. 

The topological floppy modes can also be revealed by a supercell analysis. We consider a strip of six unit cells, as displayed in Fig.~\ref{fig:Supercell_analysis}(a). We apply 1D Bloch conditions along vector $\mathbf{a}$ and keep the left and right edges free. In Fig.~\ref{fig:Supercell_analysis}(b), we report the band diagrams (up to the first four bands) for four sets of spring parameters: (i) $K_s=0.5K_l$ and $K_\theta\neq0$; (ii) $K_s=K_l$ and $K_\theta\neq0$; (iii) $K_s=0.5K_l$ and $K_\theta=0$; (iv) $K_s=K_l$ and $K_\theta=0$. The rest parameters $M_a=M_b=1$, $J_a=J_b=1$, $K_l=1$ are the same for all cases and $K_\theta=1\times10^{-3}$ if chosen to be zero. For case (i) where $K_\theta=0$, we observe two zero modes (mathematically it can be shown that $\text{rank}\left(\mathbf{K}(\mathbf{k})\right)=6n_d-2$), and the mode shapes shown in Fig.~\ref{fig:Supercell_analysis}(c) are calculated for one zero mode at selected wavenumbers indicated by red dots in Fig.~\ref{fig:Supercell_analysis}(b). As predicted by the the polarization vector $\mathbf{R}_T$, these mode shapes indeed feature displacements that are localized at the left edge. For case (ii) where $K_\theta\neq0$, the two zero modes morph into finite-frequency edge modes. Specifically, mode shapes calculated along the lowest branch are displayed in Fig.~\ref{fig:Supercell_analysis}(c), which exhibit displacement localization at the left edge especially for large wavenumbers, highly consistent with those for case (i). This observation indicates that the topological characteristics, to a large degree, are preserved for cases where $K_\theta\neq0$ (mode shapes for the second lowest branch in case (ii) are documented in Appendix C, showing similar topological characteristics). Interestingly, we note that the shear stiffness $K_s$ doesn't affect the zero modes in cases (i) and (iii) and the edge modes in cases (ii) and (iv), which may be ascribed to the fact that these topological modes intrinsically stem from the geometry of the lattice (i.e., the compatibility matrix $\mathbf{C}(\mathbf{k})$), instead of the spring constants. Finally, it is worth noting that our findings conform to those obtained previously for ideal kagome lattices and physical specimens \cite{Kane_NP_2014,Ma_PRL_2018}. 

\section{Conclusions}
In conclusion, we have established a theoretical framework, validated through finite element analysis, to characterize the dynamic properties of physically-realizable kagome metamaterials, in which triangles and ligaments are modeled as rigid bodies and elastic springs, respectively. Using the proposed framework, we have demonstrated, when certain spring parameters are carefully chosen, the dualities between the stiffness matrices of pairs of  twisted kagome metamaterials and twofold degeneracy in the self-dual kagome metamaterial. By applying the framework to deformed kagome metamaterials, we have theoretically revealed the existence of topological zero modes in cases where there is no torsional constraint between triangles (i.e., torsional stiffness $K_\theta=0$). Moreover, we have unequivocally shown that it is the nonzero torsional stiffness $K_\theta\neq0$ that causes these zero modes to morph into finite-frequency edge modes. Therefore, we have shown that the structural effects play an important role on these dynamic properties. The presented framework can be applied to other physically-realizable kagome systems as long as the interactions between triangles can be effectively modeled as elastic springs, which could unlock new design strategies to achieve novel wave manipulation functionalities in kagome-based metamaterials. 

\section*{Acknowledgments}
This work was supported by the Fundamental Research Funds for the Central Universities (grant number 22120240045) and Shanghai Gaofeng Project for University Academic Program Development. H.S. acknowledges support via NSF award number 2041410.

\section*{Appendix A}\label{Appendix A}
\setcounter{figure}{0}
\setcounter{equation}{0}
\renewcommand{\thefigure}{S\arabic{figure}}
\renewcommand{\theequation}{S\arabic{equation}}
The relative displacements $\Delta \mathbf{l}_{n,m}$ and $\Delta \mathbf{l}_{n\pm1,m\pm1}$ in Eq.~\ref{Potential_comp} can be expressed as 
\begin{equation}
\begin{split}
\Delta \mathbf{l}_{n,m}&=  \mathbf{r}_{1}^b-\mathbf{r}_{1}^a+\mathbf{u}^a_{n,m} +\mathbf{C}_{n,m}^a\mathbf{r}_{1}^a-\mathbf{u}^b_{n,m}-\mathbf{C}_{n,m}^b \mathbf{r}_{1}^b\\
\Delta \mathbf{l}_{n+1,m}&=\mathbf{e}_1 +\mathbf{u}^b_{n+1,m} +\mathbf{C}_{n+1,m}^b\mathbf{r}_{2}^b-(\mathbf{r}_{1}^b-\mathbf{r}_{1}^a+\mathbf{u}^a_{n,m} +\mathbf{C}_{n,m}^a\mathbf{r}_{2}^a),\\
\Delta \mathbf{l}_{n-1,m}&=\mathbf{r}_{1}^b-\mathbf{r}_{1}^a-\mathbf{e}_1 +\mathbf{u}^a_{n-1,m} +\mathbf{C}_{n-1,m}^a\mathbf{r}_{2}^a-\mathbf{u}^b_{n,m} -\mathbf{C}_{n,m}^b\mathbf{r}_{2}^b,\\
\Delta \mathbf{l}_{n,m+1}&=\mathbf{e}_2 +\mathbf{u}^b_{n,m+1} +\mathbf{C}_{n,m+1}^b\mathbf{r}_{3}^b-(\mathbf{r}_{1}^b-\mathbf{r}_{1}^a+\mathbf{u}^a_{n,m} +\mathbf{C}_{n,m}^a\mathbf{r}_{3}^a),\\
\Delta \mathbf{l}_{n,m-1}&=\mathbf{r}_{1}^b-\mathbf{r}_{1}^a-\mathbf{e}_2 +\mathbf{u}^a_{n,m-1} +\mathbf{C}_{n,m-1}^a\mathbf{r}_{3}^a-\mathbf{u}^b_{n,m} -\mathbf{C}_{n,m}^b\mathbf{r}_{3}^b,
\end{split}
\end{equation}
where $\mathbf{C}_{n,m}^a=\begin{bmatrix} \cos \theta^a_{n,m} & -\sin \theta^a_{n,m}\\ \sin \theta^a_{n,m} & \cos \theta^a_{n,m} \end{bmatrix}$, $\mathbf{C}_{n,m}^b=\begin{bmatrix} \cos \theta^b_{n,m} & \sin \theta^b_{n,m}\\ -\sin \theta^b_{n,m} & \cos \theta^b_{n,m} \end{bmatrix}$, $\mathbf{e}_1=\mathbf{r}_{1}^b-\mathbf{r}_{2}^b-\mathbf{r}_{1}^a+\mathbf{r}_{2}^a$, and $\mathbf{e}_2=\mathbf{r}_{1}^b-\mathbf{r}_{3}^b-\mathbf{r}_{1}^a+\mathbf{r}_{3}^a$. 

In the linear wave analysis, the mass matrix $\mathbf{M}$ 
 and the stiffness matrices $\mathbf{K}_i$ in Eq.~\ref{Govering_eqn} can be written as 
\begin{equation}
\begin{split}
\mathbf{M}&=\begin{bmatrix} M_a & 0 & 0 & 0 & 0 & 0\\ 0 & M_a & 0 & 0 & 0 & 0\\0 & 0 & M_b & 0 & 0 & 0\\0 & 0 & 0 & M_b & 0 & 0\\0 & 0 & 0 & 0 & J_a & 0\\0 & 0 & 0 & 0 & 0 & J_b\\ \end{bmatrix},
\;\;\;
\mathbf{K}_1=\begin{bmatrix} d_{11} & d_{12} & d_{13}\\ d_{21} & d_{22} & d_{23}\\d_{31} & d_{32} & d_{33} \end{bmatrix},
\;\;\;
\mathbf{K}_2=\begin{bmatrix} 0 & 0 & 0\\ e_{21} & 0 & e_{23}\\e_{31} & 0 & e^b_{33} \end{bmatrix},
\;\;\;
\mathbf{K}_3=\begin{bmatrix} 0 & e_{12} & e_{13}\\ 0 & 0 & 0\\0 & e_{32} & e^a_{33} \end{bmatrix},\\
\mathbf{K}_4&=\begin{bmatrix} 0 & 0 & 0\\ f_{21} & 0 & f_{23}\\f_{31} & 0 & f^b_{33} \end{bmatrix},
\;\;\;
\mathbf{K}_5=\begin{bmatrix} 0 & f_{12} & f_{13}\\ 0 & 0 & 0\\0 & f_{32} & f^a_{33} \end{bmatrix}.
\end{split}
\end{equation}
The components of the above matrices are given as
\begin{equation}
\begin{split}
d_{11}&=K_l \mathbf{p}_1\otimes\mathbf{p}_1+K_s\mathbf{p}_2\otimes\mathbf{p}_2+K_l \mathbf{n}_1\otimes\mathbf{n}_1+K_s\mathbf{n}_2\otimes\mathbf{n}_2+K_l \mathbf{m}_1\otimes\mathbf{m}_1+K_s\mathbf{m}_2\otimes\mathbf{m}_2,\\
d_{12}&=- (K_l \mathbf{p}_1\otimes\mathbf{p}_1+K_s\mathbf{p}_2\otimes\mathbf{p}_2),\\
d_{13}&=(K_l \mathbf{p}_1\otimes\mathbf{p}_1+K_s\mathbf{p}_2\otimes\mathbf{p}_2)\mathbf{\gamma}_1+(K_l \mathbf{n}_1\otimes\mathbf{n}_1+K_s\mathbf{n}_2\otimes\mathbf{n}_2)\left(\mathbf{\gamma}_2\odot\mathbf{\eta}_1\right)+(K_l \mathbf{m}_1\otimes\mathbf{m}_1+K_s\mathbf{m}_2\otimes\mathbf{m}_2)\left(\mathbf{\gamma}_3\odot\mathbf{\eta}_1\right),\\
d_{21}&=-(K_l \mathbf{p}_1\otimes\mathbf{p}_1+K_s\mathbf{p}_2\otimes\mathbf{p}_2),\\
d_{22}&=K_l \mathbf{p}_1\otimes\mathbf{p}_1+K_s\mathbf{p}_2\otimes\mathbf{p}_2+K_l \mathbf{n}_1\otimes\mathbf{n}_1+K_s\mathbf{n}_2\otimes\mathbf{n}_2+K_l \mathbf{m}_1\otimes\mathbf{m}_1+K_s\mathbf{m}_2\otimes\mathbf{m}_2,\\
d_{23}&=-(K_l \mathbf{p}_1\otimes\mathbf{p}_1+K_s\mathbf{p}_2\otimes\mathbf{p}_2)\mathbf{\gamma}_1-(K_l \mathbf{n}_1\otimes\mathbf{n}_1+K_s\mathbf{n}_2\otimes\mathbf{n}_2)\left(\mathbf{\gamma}_2\odot\mathbf{\eta}_2\right)-(K_l \mathbf{m}_1\otimes\mathbf{m}_1+K_s\mathbf{m}_2\otimes\mathbf{m}_2)\left(\mathbf{\gamma}_3\odot\mathbf{\eta}_2\right),\\
d_{31}&=\begin{bmatrix}\left( K_l\mathbf{p}_1^T\mathbf{C'}^{a}\mathbf{r}_{1}^a\right)\mathbf{p}^T_1+\left(K_s\mathbf{p}_2^T\mathbf{C'}^{a}\mathbf{r}_{1}^a\right)\mathbf{p}^T_2\\-\left( K_l\mathbf{p}_1^T\mathbf{C'}^{b}\mathbf{r}_{1}^b\right)\mathbf{p}^T_1-\left( K_s\mathbf{p}_2^T\mathbf{C'}^{b}\mathbf{r}_{1}^b\right)\mathbf{p}^T_2 \end{bmatrix}+\begin{bmatrix}\left( K_l\mathbf{n}_1^T\mathbf{C'}^{a}\mathbf{r}_{2}^a\right)\mathbf{n}^T_1+\left(K_s\mathbf{n}_2^T\mathbf{C'}^{a}\mathbf{r}_{2}^a\right)\mathbf{n}^T_2\\ (0\,\,\, 0) \end{bmatrix}\\
&+\begin{bmatrix}\left( K_l\mathbf{m}_1^T\mathbf{C'}^{a}\mathbf{r}_{3}^a\right)\mathbf{m}^T_1+\left(K_s\mathbf{m}_2^T\mathbf{C'}^{a}\mathbf{r}_{3}^a\right)\mathbf{m}^T_2\\ (0\,\,\, 0) \end{bmatrix},\\
d_{32}&=-\begin{bmatrix}\left( K_l\mathbf{p}_1^T\mathbf{C'}^{a}\mathbf{r}_{1}^a\right)\mathbf{p}^T_1+\left(K_s\mathbf{p}_2^T\mathbf{C'}^{a}\mathbf{r}_{1}^a\right)\mathbf{p}^T_2\\-\left( K_l\mathbf{p}_1^T\mathbf{C'}^{b}\mathbf{r}_{1}^b\right)\mathbf{p}^T_1-\left( K_s\mathbf{p}_2^T\mathbf{C'}^{b}\mathbf{r}_{1}^b\right)\mathbf{p}^T_2 \end{bmatrix}+\begin{bmatrix}(0\,\,\, 0) \\ \left( K_l\mathbf{n}_1^T\mathbf{C'}^{b}\mathbf{r}_{2}^b\right)\mathbf{n}^T_1+\left(K_s\mathbf{n}_2^T\mathbf{C'}^{b}\mathbf{r}_{2}^b\right)\mathbf{n}^T_2  \end{bmatrix}\\
&+\begin{bmatrix}(0\,\,\, 0) \\ \left( K_l\mathbf{m}_1^T\mathbf{C'}^{b}\mathbf{r}_{3}^b\right)\mathbf{m}^T_1+\left(K_s\mathbf{m}_2^T\mathbf{C'}^{b}\mathbf{r}_{3}^b\right)\mathbf{m}^T_2\end{bmatrix},\\
d_{33}&=K_\theta\mathbf{\zeta}+\begin{bmatrix}\left( K_l\mathbf{p}_1^T\mathbf{C'}^{a}\mathbf{r}_{1}^a\right)\mathbf{p}^T_1+\left(K_s\mathbf{p}_2^T\mathbf{C'}^{a}\mathbf{r}_{1}^a\right)\mathbf{p}^T_2\\-\left( K_l\mathbf{p}_1^T\mathbf{C'}^{b}\mathbf{r}_{1}^b\right)\mathbf{p}^T_1-\left( K_s\mathbf{p}_2^T\mathbf{C'}^{b}\mathbf{r}_{1}^b\right)\mathbf{p}^T_2 \end{bmatrix}\mathbf{\gamma}_1+\left(\begin{bmatrix}\left( K_l\mathbf{n}_1^T\mathbf{C'}^{a}\mathbf{r}_{2}^a\right)\mathbf{n}^T_1+\left(K_s\mathbf{n}_2^T\mathbf{C'}^{a}\mathbf{r}_{2}^a\right)\mathbf{n}^T_2\\ -\left( K_l\mathbf{n}_1^T\mathbf{C'}^{b}\mathbf{r}_{2}^b\right)\mathbf{n}^T_1-\left(K_s\mathbf{n}_2^T\mathbf{C'}^{b}\mathbf{r}_{2}^b\right)\mathbf{n}^T_2 \end{bmatrix}\mathbf{\gamma}_2\right)\odot\mathbf{\eta}_3\\
&+\left(\begin{bmatrix}\left( K_l\mathbf{m}_1^T\mathbf{C'}^{a}\mathbf{r}_{3}^a\right)\mathbf{m}^T_1+\left(K_s\mathbf{m}_2^T\mathbf{C'}^{a}\mathbf{r}_{3}^a\right)\mathbf{m}^T_2\\ -\left( K_l\mathbf{m}_1^T\mathbf{C'}^{b}\mathbf{r}_{3}^b\right)\mathbf{m}^T_1-\left(K_s\mathbf{m}_2^T\mathbf{C'}^{b}\mathbf{r}_{3}^b\right)\mathbf{m}^T_2 \end{bmatrix}\mathbf{\gamma}_3\right)\odot\mathbf{\eta}_3,
\end{split}
\end{equation}
\begin{equation}
\begin{split}
e_{21}&=-(K_l \mathbf{n}_1\otimes\mathbf{n}_1+K_s\mathbf{n}_2\otimes\mathbf{n}_2),\\
e_{23}&=-(K_l \mathbf{n}_1\otimes\mathbf{n}_1+K_s\mathbf{n}_2\otimes\mathbf{n}_2)\left(\mathbf{\gamma}_2\odot\eta_1\right),\\
e_{31}&=-\begin{bmatrix}(0\,\,\, 0) \\ \left( K_l\mathbf{n}_1^T\mathbf{C'}^{b}\mathbf{r}_{2}^b\right)\mathbf{n}^T_1+\left(K_s\mathbf{n}_2^T\mathbf{C'}^{b}\mathbf{r}_{2}^b\right)\mathbf{n}^T_2  \end{bmatrix},\\
e_{33}^b&=\begin{bmatrix} 0 & 0\\ K_\theta & 0 \end{bmatrix}-\begin{bmatrix}(0\,\,\, 0) \\ \left( K_l\mathbf{n}_1^T\mathbf{C'}^{b}\mathbf{r}_{2}^b\right)\mathbf{n}^T_1+\left(K_s\mathbf{n}_2^T\mathbf{C'}^{b}\mathbf{r}_{2}^b\right)\mathbf{n}^T_2 \end{bmatrix}\left(\mathbf{\gamma}_2\odot\eta_1\right),\\
e_{12}&=-(K_l \mathbf{n}_1\otimes\mathbf{n}_1+K_s\mathbf{n}_2\otimes\mathbf{n}_2),\\
e_{13}&=(K_l \mathbf{n}_1\otimes\mathbf{n}_1+K_s\mathbf{n}_2\otimes\mathbf{n}_2)\left(\mathbf{\gamma}_2\odot\eta_2\right),\\
e_{32}&=-\begin{bmatrix} \left( K_l\mathbf{n}_1^T\mathbf{C'}^{a}\mathbf{r}_{2}^a\right)\mathbf{n}^T_1+\left(K_s\mathbf{n}_2^T\mathbf{C'}^{a}\mathbf{r}_{2}^a\right)\mathbf{n}^T_2\\ (0\,\,\, 0) \end{bmatrix},\\
e_{33}^a&=\begin{bmatrix} 0 & K_\theta\\ 0 & 0 \end{bmatrix}+\begin{bmatrix} \left( K_l\mathbf{n}_1^T\mathbf{C'}^{a}\mathbf{r}_{2}^a\right)\mathbf{n}^T_1+\left(K_s\mathbf{n}_2^T\mathbf{C'}^{a}\mathbf{r}_{2}^a\right)\mathbf{n}^T_2\\ (0\,\,\, 0) \end{bmatrix}\left(\mathbf{\gamma}_2\odot\eta_2\right),\\
f_{21}&=-(K_l \mathbf{m}_1\otimes\mathbf{m}_1+K_s\mathbf{m}_2\otimes\mathbf{m}_2),\\
f_{23}&=(K_l \mathbf{m}_1\otimes\mathbf{m}_1+K_s\mathbf{m}_2\otimes\mathbf{m}_2)\left(\mathbf{\gamma}_3\odot\eta_1\right),\\
f_{31}&=-\begin{bmatrix}(0\,\,\, 0) \\ \left( K_l\mathbf{m}_1^T\mathbf{C'}^{b}\mathbf{r}_{3}^b\right)\mathbf{m}^T_1+\left(K_s\mathbf{m}_2^T\mathbf{C'}^{b}\mathbf{r}_{3}^b\right)\mathbf{m}^T_2  \end{bmatrix},\\
f_{33}^b&=\begin{bmatrix} 0 & 0\\ K_\theta & 0 \end{bmatrix}-\begin{bmatrix}(0\,\,\, 0) \\ \left( K_l\mathbf{m}_1^T\mathbf{C'}^{b}\mathbf{r}_{3}^b\right)\mathbf{m}^T_1+\left(K_s\mathbf{m}_2^T\mathbf{C'}^{b}\mathbf{r}_{3}^b\right)\mathbf{m}^T_2  \end{bmatrix}\left(\mathbf{\gamma}_3\odot\eta_1\right),\\
f_{12}&=-(K_l \mathbf{m}_1\otimes\mathbf{m}_1+K_s\mathbf{m}_2\otimes\mathbf{m}_2),\\
f_{13}&=(K_l \mathbf{m}_1\otimes\mathbf{m}_1+K_s\mathbf{m}_2\otimes\mathbf{m}_2)\left(\mathbf{\gamma}_3\odot\eta_2\right),\\
f_{32}&=-\begin{bmatrix} \left( K_l\mathbf{m}_1^T\mathbf{C'}^{a}\mathbf{r}_{3}^a\right)\mathbf{m}^T_1+\left(K_s\mathbf{m}_2^T\mathbf{C'}^{a}\mathbf{r}_{3}^a\right)\mathbf{m}^T_2\\ (0\,\,\, 0) \end{bmatrix},\\
f_{33}^a&=\begin{bmatrix} 0 & K_\theta\\ 0 & 0 \end{bmatrix}+\begin{bmatrix} \left( K_l\mathbf{m}_1^T\mathbf{C'}^{a}\mathbf{r}_{3}^a\right)\mathbf{m}^T_1+\left(K_s\mathbf{m}_2^T\mathbf{C'}^{a}\mathbf{r}_{3}^a\right)\mathbf{m}^T_2\\ (0\,\,\, 0) \end{bmatrix}\left(\mathbf{\gamma}_3\odot\eta_2\right),\\
\end{split}
\end{equation}
where $\odot$ denotes element-wise multiplication,
$\mathbf{\gamma}_i=\begin{bmatrix} -\mathbf{r}_{i}^a(2) & -\mathbf{r}_{i}^b(2)\\ \mathbf{r}_{i}^a(1) & \mathbf{r}_{i}^b(1) \end{bmatrix}$ with $i\in[1,2,3]$, $\mathbf{C'}^{a}=\begin{bmatrix} 0 & -1\\ 1 & 0 \end{bmatrix}$, $\mathbf{C'}^{b}=\begin{bmatrix} 0 & 1\\ -1 & 0 \end{bmatrix}$, $\mathbf{\eta}_1=\begin{bmatrix} 1 & 0\\ 1 & 0 \end{bmatrix}$, $\mathbf{\eta}_2=\begin{bmatrix} 0 & 1\\ 0 & 1 \end{bmatrix}$, $\mathbf{\eta}_3=\begin{bmatrix} 1 & 0\\ 0 & 1 \end{bmatrix}$, and $\mathbf{\zeta}=\begin{bmatrix} 3 & 1\\ 1 & 3 \end{bmatrix}$.

Then, the global stiffness matrix $\mathbf{K}$ can be expressed as
\begin{equation}
\begin{split}
\mathbf{K}&=\begin{bmatrix} d_{11} & d_{12} & d_{13}\\ d_{21} & d_{22} & d_{23}\\d_{31} & d_{32} & d_{33} \end{bmatrix}+\begin{bmatrix} 0 & e_{12}e^{i\mathbf{k} \cdot \mathbf{e}_1} +f_{12}e^{i\mathbf{k} \cdot \mathbf{e}_2}& e_{13}e^{i\mathbf{k} \cdot \mathbf{e}_1}+f_{13}e^{i\mathbf{k} \cdot \mathbf{e}_2}\\ e_{21}e^{-i\mathbf{k} \cdot \mathbf{e}_1}+f_{21}e^{-i\mathbf{k} \cdot \mathbf{e}_2}& 0 & e_{23}e^{-i\mathbf{k} \cdot \mathbf{e}_1}+f_{23}e^{-i\mathbf{k} \cdot \mathbf{e}_2}\\e_{31}e^{-i\mathbf{k} \cdot \mathbf{e}_1}+f_{31}e^{-i\mathbf{k} \cdot \mathbf{e}_2} & e_{32}e^{i\mathbf{k} \cdot \mathbf{e}_1} +f_{32}e^{i\mathbf{k} \cdot \mathbf{e}_2}& e^b_{33}e^{-i\mathbf{k} \cdot \mathbf{e}_1}+e^a_{33}e^{i\mathbf{k} \cdot \mathbf{e}_1}+f^b_{33}e^{-i\mathbf{k} \cdot \mathbf{e}_2} +f^a_{33}e^{i\mathbf{k} \cdot \mathbf{e}_2}\end{bmatrix}\\
&\equiv\begin{bmatrix} k_{11} & k_{12} & k_{13}\\ k_{21} & k_{22} & k_{23}\\k_{31} & k_{32} & k_{33} \end{bmatrix}.
\end{split}
\end{equation}
\clearpage
\section*{Appendix B}\label{Appendix B}

\begin{figure*}[htbp]
    \centerline{ \includegraphics[width=0.7\textwidth]{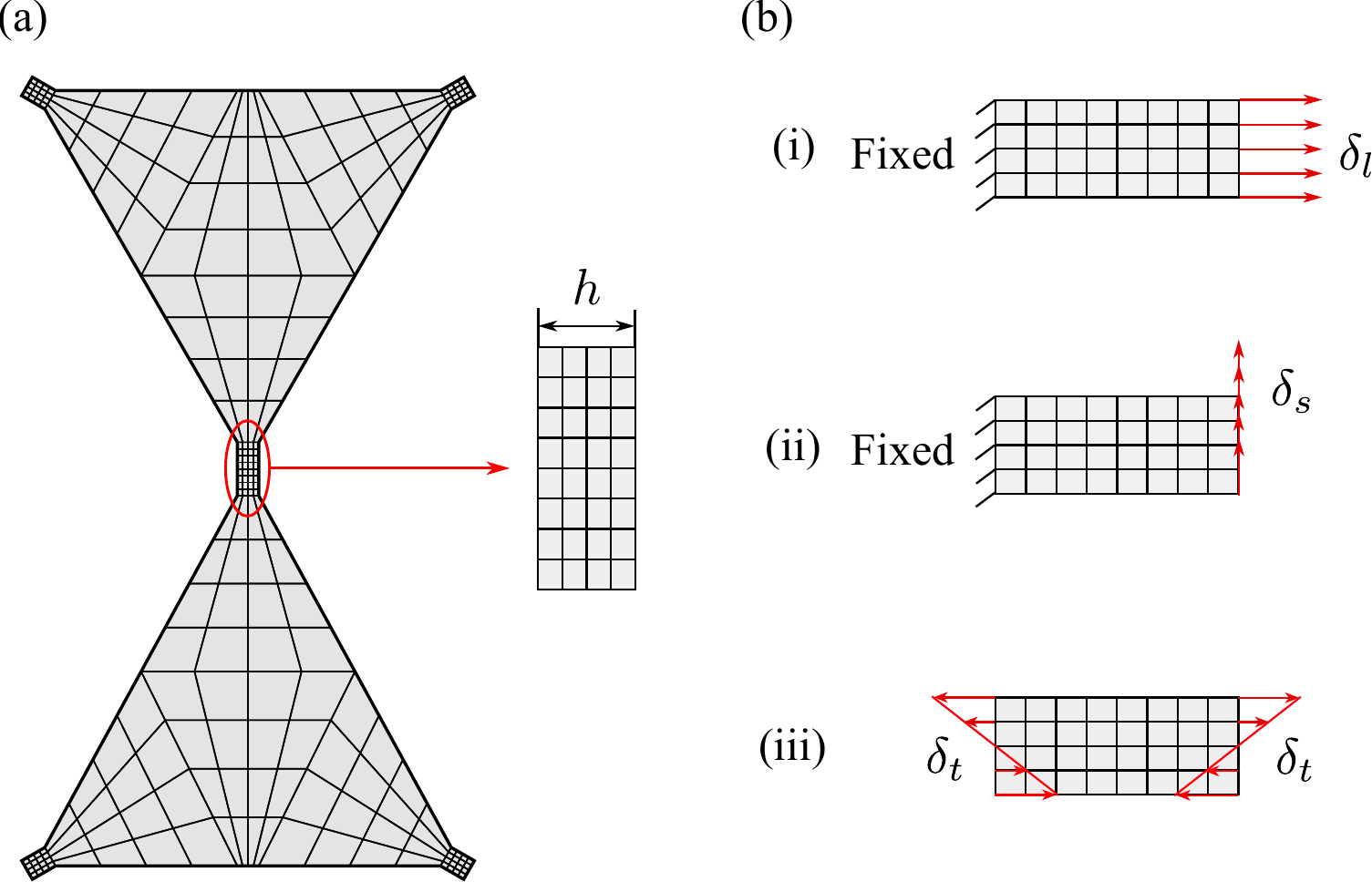}}
    \caption{Finite element models for the unit cell and the ligament of a regular kagome metamaterial. (a) Meshing of the unit cell with an inset highlighting the ligament. (b) Schematics of the boundary conditions in the FE simulations to determine the equivalent (i) longitudinal, (ii) shear, and (iii) torsional stiffnesses of the ligament.} 
    \label{fig:Finite_element_model}
\end{figure*}
In the FE analysis of the regular kagome metamterial studied in Section III, the unit cell of the metamaterial is discretized using plane-stress elements. Fig.~\ref{fig:Finite_element_model}(a) shows the meshing of the unit cell with the ligament (with width $h$) highlighted in the inset. Using the same meshing properties, we perform a series of FE simulations of a single ligament to determine its equivalent spring stiffnesses. As shown in Fig.~\ref{fig:Finite_element_model}(b), we consider three boundary conditions:
(i) the left boundary is fixed and a horizontal displacement $\delta_l$ is applied to the nodes on the right boundary; (ii) the left boundary is fixed and a vertical displacement $\delta_s$ is applied to the nodes on the right boundary; (iii) a displacement profile $\delta_t$ is applied to the nodes on the left and right boundaries.

Then, the equivalent stiffnesses $K_l$, $K_s$ and $K_\theta$ are determined as
\begin{equation}
K_l=\frac{F_l}{\delta_l}, \,\,\, K_s=\frac{F_s}{\delta_s}, \,\,\, K_l=\frac{M_th}{4\max(\delta_l)},
\end{equation}
where $F_l$ and $F_s$ are the horizontal and vertical reaction forces, respectively, calculated by summing all reaction forces at the nodes on the right boundary, and $M_t$ is the reaction moment obtained by summing the moment generated by every nodal force on the right boundary.
\clearpage
\section*{Appendix C}\label{Appendix C}
\begin{figure*}[htbp]
    \centerline{ \includegraphics[width=1\textwidth]{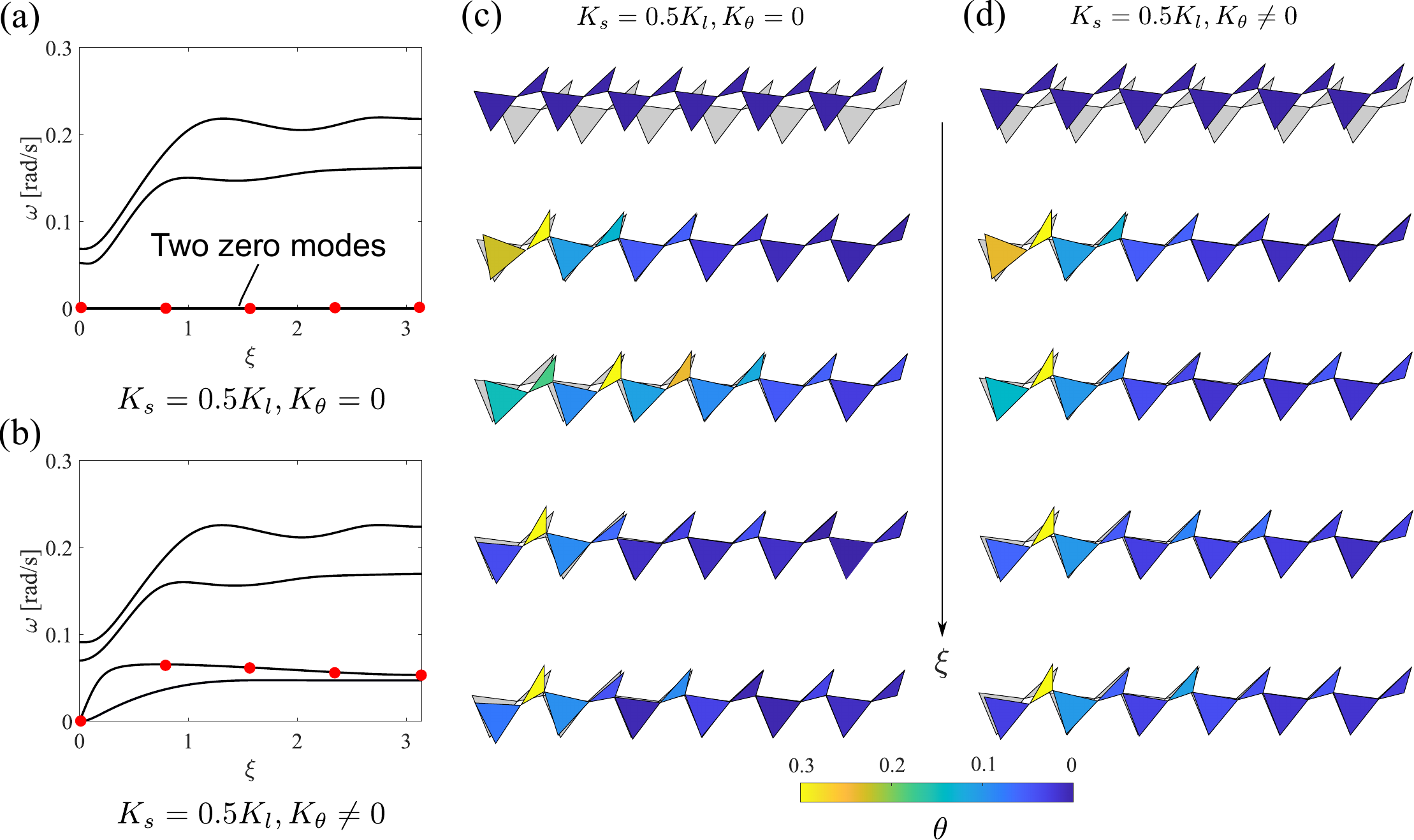}}
    \caption{Topological edge modes in deformed kagome metamaterials. (a-b) Band diagram of the supercell for (a) $K_s=0.5K_l$ and $K_\theta=0$, and (b) $K_s=0.5K_l$ and $K_\theta\neq0$. (c-d) Mode shapes calculated along the other zero mode in case (a), i.e., case (i) in Fig.4(b), and along the second lowest edge mode in case (b), i.e., case (ii) in Fig.4(b), at selected wavenumbers indicated by red dots in (a-b).} 
    \label{fig:Supercell_analysis_SI}
\end{figure*}

In Fig.~\ref{fig:Supercell_analysis_SI}(c), we show mode shapes calculated at selected wavenumbers for the other zero mode in case (i) of Fig.4(b), replotted in Fig.~\ref{fig:Supercell_analysis_SI}(a). For comparison, we in Fig.~\ref{fig:Supercell_analysis_SI}(d) show mode shapes calculated at selected wavenumbers for the second lowest edge mode in case (ii) of Fig.4(b), replotted in Fig.~\ref{fig:Supercell_analysis_SI}(b). We observe that these mode shapes feature displacements that are highly localized at the left edge, qualitatively consistent with the topological characteristics shown in Fig.4(c-d). 

\bibliographystyle{unsrt}
\bibliography{ref}

\begin{thebibliography}{10}

\bibitem{Hussein_2014}
M.I. Hussein, M.J. Leamy, and M.~Ruzzene.
\newblock Dynamics of phononic materials and structures: Historical origins, recent progress, and future outlook.
\newblock {\em Applied Mechanics Reviews}, 66(4):040802, 2014.

\bibitem{bertoldi2017flexible}
Katia Bertoldi, Vincenzo Vitelli, Johan Christensen, and Martin Van~Hecke.
\newblock Flexible mechanical metamaterials.
\newblock {\em Nature Reviews Materials}, 2(11):1--11, 2017.

\bibitem{fronk2023_ND}
Matthew~D Fronk, Lezheng Fang, Pawel Packo, and Michael~J Leamy.
\newblock Elastic wave propagation in weakly nonlinear media and metamaterials: a review of recent developments.
\newblock {\em Nonlinear Dynamics}, 111(12):10709--10741, 2023.

\bibitem{liu2000locally}
Zhengyou Liu, Xixiang Zhang, Yiwei Mao, YY~Zhu, Zhiyu Yang, Che~Ting Chan, and Ping Sheng.
\newblock Locally resonant sonic materials.
\newblock {\em science}, 289(5485):1734--1736, 2000.

\bibitem{tournat_njp_2011}
Vincent Tournat, Isabel P{\`e}rez-Arjona, Aur{\'e}lien Merkel, Victor Sanchez-Morcillo, and Vitalyi Gusev.
\newblock Elastic waves in phononic monolayer granular membranes.
\newblock {\em New Journal of Physics}, 13(7):073042, 2011.

\bibitem{Tournat_PRL_2011}
A.~Merkel, V.~Tournat, and V.~Gusev.
\newblock Experimental evidence of rotational elastic waves in granular phononic crystals.
\newblock {\em Phys. Rev. Lett.}, 107:225502, Nov 2011.

\bibitem{Tournat_pre_2014}
H.~Pichard, A.~Duclos, J.-P. Groby, V.~Tournat, and V.~E. Gusev.
\newblock Localized transversal-rotational modes in linear chains of equal masses.
\newblock {\em Phys. Rev. E}, 89:013201, Jan 2014.

\bibitem{Tournat_APL_2016}
F~Allein, V~Tournat, VE~Gusev, and Georgios Theocharis.
\newblock Tunable magneto-granular phononic crystals.
\newblock {\em Applied Physics Letters}, 108(16), 2016.

\bibitem{Tournat_PRE_2016}
H.~Pichard, A.~Duclos, J.-P. Groby, V.~Tournat, L.~Zheng, and V.~E. Gusev.
\newblock Surface waves in granular phononic crystals.
\newblock {\em Phys. Rev. E}, 93:023008, Feb 2016.

\bibitem{Tournat_EML_2017}
Li-Yang Zheng, Vincent Tournat, and Vitalyi Gusev.
\newblock Zero-frequency and extremely slow elastic edge waves in mechanical granular graphene.
\newblock {\em Extreme Mechanics Letters}, 12:55--64, 2017.

\bibitem{Tournat_EML_2017b}
F~Allein, V~Tournat, VE~Gusev, and Georgios Theocharis.
\newblock Transversal--rotational and zero group velocity modes in tunable magneto-granular phononic crystals.
\newblock {\em Extreme Mechanics Letters}, 12:65--70, 2017.

\bibitem{Liang_2010}
B.~Liang, X.S. Guo, J.~Tu, D.~Zhang, and J.~C. Cheng.
\newblock An acoustic rectifier.
\newblock {\em Nature Materials}, 9:989, 2010.

\bibitem{Boechler_2011}
N.~Boechler, G.~Theocharis, and C.~Daraio.
\newblock Bifurcation-based acoustic switching and rectification.
\newblock {\em Nature Materials}, 10(9):665 -- 668, 2011.

\bibitem{Bilal_2017}
O.~R. Bilal, A.~Foehr, and C.~Daraio.
\newblock Bistable metamaterial for switching and cascading elastic vibrations.
\newblock {\em Proc. Nat. Acad. Sci. USA}, 114(18):4603--4606, 2017.

\bibitem{Zhou_PRB2020}
Di~Zhou, Jihong Ma, Kai Sun, Stefano Gonella, and Xiaoming Mao.
\newblock Switchable phonon diodes using nonlinear topological maxwell lattices.
\newblock {\em Physical Review B}, 101(10):104106, 2020.

\bibitem{Ganesh_2017}
R.~Ganesh and S.~Gonella.
\newblock Nonlinear waves in lattice materials: Adaptively augmented directivity and functionality enhancement by modal mixing.
\newblock {\em Journal of the Mechanics and Physics of Solids}, 99:272 -- 288, 2017.

\bibitem{JIAO_JMPS_2018}
W.~Jiao and S.~Gonella.
\newblock Mechanics of inter-modal tunneling in nonlinear waveguides.
\newblock {\em Journal of the Mechanics and Physics of Solids}, 111:1--17, 2018.

\bibitem{Daraio_2006}
C.~Daraio, V.~F. Nesterenko, E.~B. Herbold, and S.~Jin.
\newblock Tunability of solitary wave properties in one-dimensional strongly nonlinear phononic crystals.
\newblock {\em Phys. Rev. E}, 73:026610, Feb 2006.

\bibitem{Narisetti_2011}
R.K. Narisetti, M.~Ruzzene, and M.J. Leamy.
\newblock A perturbation approach for analyzing dispersion and group velocities in two-dimensional nonlinear periodic lattices.
\newblock {\em Journal of Vibration and Acoustics}, 133(6):061020, 2011.

\bibitem{Jiao_prsa_2021}
Weijian Jiao and Stefano Gonella.
\newblock Wavenumber-space band clipping in nonlinear periodic structures.
\newblock {\em Proceedings of the Royal Society A: Mathematical, Physical and Engineering Sciences}, 477(2251):20210052, 2021.

\bibitem{Zhu_NC_2014}
R.~Zhu, X.~N. Liu, G.~K. Hu, C.~T. Sun, and G.~L. Huang.
\newblock Negative refraction of elastic waves at the deep-subwavelength scale in a single-phase metamaterial.
\newblock {\em Nat. Commun.}, 5:5510, 2014.

\bibitem{Raney2016}
J.~R. Raney, N.~Nadkarni, C.~Daraio, D.~M. Kochmann, J.~A. Lewis, and K.~Bertoldi.
\newblock {Stable propagation of mechanical signals in soft media using stored elastic energy}.
\newblock {\em Proc. Nat. Acad. Sci. USA}, 113(35):9722--9727, 2016.

\bibitem{Nadkarni_PRL_2016}
Neel Nadkarni, Andres~F. Arrieta, Christopher Chong, Dennis~M. Kochmann, and Chiara Daraio.
\newblock Unidirectional transition waves in bistable lattices.
\newblock {\em Phys. Rev. Lett.}, 116:244501, Jun 2016.

\bibitem{Wang_PRL_2018}
Yifan Wang, Behrooz Yousefzadeh, Hui Chen, Hussein Nassar, Guoliang Huang, and Chiara Daraio.
\newblock Observation of nonreciprocal wave propagation in a dynamic phononic lattice.
\newblock {\em Phys. Rev. Lett.}, 121:194301, Nov 2018.

\bibitem{Moore_PRE_2018}
Keegan~J. Moore, Jonathan Bunyan, Sameh Tawfick, Oleg~V. Gendelman, Shuangbao Li, Michael Leamy, and Alexander~F. Vakakis.
\newblock Nonreciprocity in the dynamics of coupled oscillators with nonlinearity, asymmetry, and scale hierarchy.
\newblock {\em Phys. Rev. E}, 97:012219, Jan 2018.

\bibitem{Schaeffer2015a}
M.~Schaeffer and M.~Ruzzene.
\newblock {Wave propagation in reconfigurable magneto-elastic kagome lattice structures}.
\newblock {\em J. Appl. Phys.}, 117(19):194903, 2015.

\bibitem{paulose_2015PNAS}
Jayson Paulose, Anne~S Meeussen, and Vincenzo Vitelli.
\newblock Selective buckling via states of self-stress in topological metamaterials.
\newblock {\em Proceedings of the National Academy of Sciences}, 112(25):7639--7644, 2015.

\bibitem{Ma_PRL_2018}
Jihong Ma, Di~Zhou, Kai Sun, Xiaoming Mao, and Stefano Gonella.
\newblock Edge modes and asymmetric wave transport in topological lattices: Experimental characterization at finite frequencies.
\newblock {\em Phys. Rev. Lett.}, 121:094301, Aug 2018.

\bibitem{Pishvar_PRApplied_2020}
Maya Pishvar and Ryan~L. Harne.
\newblock Soft topological metamaterials with pronounced polar elasticity in mechanical and dynamic behaviors.
\newblock {\em Phys. Rev. Appl.}, 14:044034, Oct 2020.

\bibitem{Charara_PRApplied_2021}
Mohammad Charara, Kai Sun, Xiaoming Mao, and Stefano Gonella.
\newblock Topological flexural modes in polarized bilayer lattices.
\newblock {\em Phys. Rev. Appl.}, 16:064011, Dec 2021.

\bibitem{Xiu_PNAS_2022}
Haning Xiu, Harry Liu, Andrea Poli, Guangchao Wan, Kai Sun, Ellen~M. Arruda, Xiaoming Mao, and Zi~Chen.
\newblock Topological transformability and reprogrammability of multistable mechanical metamaterials.
\newblock {\em Proceedings of the National Academy of Sciences}, 119(52):e2211725119, 2022.

\bibitem{charara2022PNAS}
Mohammad Charara, James McInerney, Kai Sun, Xiaoming Mao, and Stefano Gonella.
\newblock Omnimodal topological polarization of bilayer networks: Analysis in the maxwell limit and experiments on a 3d-printed prototype.
\newblock {\em Proceedings of the National Academy of Sciences}, 119(40):e2208051119, 2022.

\bibitem{Zhang_PRApplied_2023}
Zi-Dong Zhang, Ming-Hui Lu, and Yan-Feng Chen.
\newblock Twist-angle-induced boundary-obstructed topological insulator on elastic kagome metamaterials.
\newblock {\em Phys. Rev. Appl.}, 20:054002, Nov 2023.

\bibitem{Azizi_PRL_2023}
Pegah Azizi, Siddhartha Sarkar, Kai Sun, and Stefano Gonella.
\newblock Dynamics of self-dual kagome metamaterials and the emergence of fragile topology.
\newblock {\em Phys. Rev. Lett.}, 130:156101, Apr 2023.

\bibitem{Li_JMPS2024}
Jian Li, Ronghao Bao, and Weiqiu Chen.
\newblock Exploring static responses, mode transitions, and feasible tunability of kagome-based flexible mechanical metamaterials.
\newblock {\em Journal of the Mechanics and Physics of Solids}, 186:105599, 2024.

\bibitem{Mao_Review_2018}
Xiaoming Mao and Tom~C. Lubensky.
\newblock Maxwell lattices and topological mechanics.
\newblock {\em Annual Review of Condensed Matter Physics}, 9(1):413--433, 2018.

\bibitem{Sun_PNAS_2012}
Kai Sun, Anton Souslov, Xiaoming Mao, and T.~C. Lubensky.
\newblock Surface phonons, elastic response, and conformal invariance in twisted kagome lattices.
\newblock {\em Proceedings of the National Academy of Sciences}, 109(31):12369--12374, 2012.

\bibitem{Kane_NP_2014}
C.~L. Kane and T.~C. Lubensky.
\newblock Topological boundary modes in isostatic lattice.
\newblock {\em Nature Phys.}, 10:39--45, 2014.

\bibitem{Danawe_PRB_2021}
Hrishikesh Danawe, Heqiu Li, Hasan~Al Ba'ba'a, and Serife Tol.
\newblock Existence of corner modes in elastic twisted kagome lattices.
\newblock {\em Phys. Rev. B}, 104:L241107, Dec 2021.

\bibitem{Fruchart_Nature_2020}
Michel Fruchart, Yujie Zhou, and Vincenzo Vitelli.
\newblock Topological boundary modes in isostatic lattice.
\newblock {\em Nature}, 577:636--640, 2020.

\bibitem{Lei_PRL_2022}
Qun-Li Lei, Feng Tang, Ji-Dong Hu, Yu-qiang Ma, and Ran Ni.
\newblock Duality, hidden symmetry, and dynamic isomerism in 2d hinge structures.
\newblock {\em Phys. Rev. Lett.}, 129:125501, Sep 2022.

\bibitem{Gonella_PRB_2020}
Stefano Gonella.
\newblock Symmetry of the phononic landscape of twisted kagome lattices across the duality boundary.
\newblock {\em Phys. Rev. B}, 102:140301, Oct 2020.

\bibitem{Stenull_PRL_2019}
Olaf Stenull and T.~C. Lubensky.
\newblock Signatures of topological phonons in superisostatic lattices.
\newblock {\em Phys. Rev. Lett.}, 122:248002, Jun 2019.

\bibitem{Nassar_JMPS_2020}
Hussein Nassar, Hui Chen, and Guoliang Huang.
\newblock Microtwist elasticity: A continuum approach to zero modes and topological polarization in kagome lattices.
\newblock {\em Journal of the Mechanics and Physics of Solids}, 144:104107, 2020.

\bibitem{Sun_PRL_2020}
Kai Sun and Xiaoming Mao.
\newblock Continuum theory for topological edge soft modes.
\newblock {\em Phys. Rev. Lett.}, 124:207601, May 2020.

\bibitem{Saremi_PRX_2020}
Adrien Saremi and Zeb Rocklin.
\newblock Topological elasticity of flexible structures.
\newblock {\em Phys. Rev. X}, 10:011052, Mar 2020.

\bibitem{Deng2017}
B.~Deng, J.~R. Raney, V.~Tournat, and K.~Bertoldi.
\newblock {Elastic Vector Solitons in Soft Architected Materials}.
\newblock {\em Phys. Rev. Lett.}, 118(20):204102, 2017.

\bibitem{Deng2019}
B.~Deng, C.~Mo, V.~Tournat, K.~Bertoldi, and J.~R. Raney.
\newblock {Focusing and mode separation of elastic vector solitons in a 2D soft mechanical metamaterial}.
\newblock {\em Phys. Rev. Lett.}, 123(2):24101, 2019.

\bibitem{ZHANG_IJSS_2023}
Quan Zhang, Andrei~V. Cherkasov, Chen Xie, Nitesh Arora, and Stephan Rudykh.
\newblock Nonlinear elastic vector solitons in hard-magnetic soft mechanical metamaterials.
\newblock {\em International Journal of Solids and Structures}, 280:112396, 2023.

\bibitem{Korpas2021}
L.~M. Korpas, R.~Yin, H.~Yasuda, and J.~R. Raney.
\newblock {Temperature-Responsive Multistable Metamaterials}.
\newblock {\em ACS Appl. Mater. Interfaces}, 13(26):31163--31170, 2021.

\bibitem{Deng_nc2018}
B.~Deng, P.~Wang, Q.~He, and K.~Bertoldi.
\newblock Metamaterials with amplitude gaps for elastic solitons.
\newblock {\em Nat. Commun.}, 9:3410, April 2018.

\bibitem{Yasuda2020}
H.~Yasuda, L.~M. Korpas, and J.~R. Raney.
\newblock {Transition Waves and Formation of Domain Walls in Multistable Mechanical Metamaterials}.
\newblock {\em Phys. Rev. Appl.}, 13(5):054067, 2020.

\bibitem{Hiromi_APL_2023}
H.~Yasuda, H.~Shu, W.~Jiao, V.~Tournat, and J.~R. Raney.
\newblock {Nucleation of transition waves via collisions of elastic vector solitons}.
\newblock {\em Applied Physics Letters}, 123(5):051701, 07 2023.

\bibitem{Jiao_NC_2024}
W.~Jiao, H.~Shu, V.~Tournat, H.~Yasuda, and J.~R. Raney.
\newblock Phase transitions in 2d multistable mechanical metamaterials via collisions of soliton-like pulses.
\newblock {\em Nat. Commun.}, 15:333, 2024.

\bibitem{Zhou_JMPS_2021}
Yuan Zhou, Yafei Zhang, and C.Q. Chen.
\newblock Amplitude-dependent boundary modes in topological mechanical lattices.
\newblock {\em Journal of the Mechanics and Physics of Solids}, 153:104482, 2021.

\bibitem{Li_IJIE_2021}
Jian Li, Yi~Yuan, Jiao Wang, Ronghao Bao, and Weiqiu Chen.
\newblock Propagation of nonlinear waves in graded flexible metamaterials.
\newblock {\em International Journal of Impact Engineering}, 156:103924, 2021.

\bibitem{CHEN_IJSS_2022}
Hui Chen, Shaoyun Wang, Xiaopeng Li, and Guoliang Huang.
\newblock Two-dimensional microtwist modeling of topological polarization in hinged kagome lattices and its experimental validation.
\newblock {\em International Journal of Solids and Structures}, 254-255:111891, 2022.

\bibitem{GUEST_JMPS_2003}
S.D Guest and J.W Hutchinson.
\newblock On the determinacy of repetitive structures.
\newblock {\em Journal of the Mechanics and Physics of Solids}, 51(3):383--391, 2003.

\bibitem{LI_JMPS_2023}
Xuenan Li and Robert~V. Kohn.
\newblock Some results on the guest–hutchinson modes and periodic mechanisms of the kagome lattice metamaterial.
\newblock {\em Journal of the Mechanics and Physics of Solids}, 178:105311, 2023.

\end{thebibliography}
\end{document}


	\title{An Analytical Model for Continuum Kagome Metamaterials} 
	
\author{Weijian Jiao}
\affiliation{School of Aerospace Engineering and Applied Mechanics, Tongji University, Shanghai 200092, China}

\author{Hang Shu}
\affiliation{Department of Mechanical Engineering and Applied Mechanics, University of Pennsylvania, Philadelphia, Pennsylvania 19104, USA}

\author{Jordan R. Raney}
\affiliation{Department of Mechanical Engineering and Applied Mechanics, University of Pennsylvania, Philadelphia, Pennsylvania 19104, USA}

	\maketitle
	\section{Equations of motion}
 Based on the discrete model introduced in the main text, the Hamiltonian of a 2D rotating-squares system can be written as
 	\begin{align}\label{Hamiltonian}
	\begin{split}
	H&=\frac{1}{2} \sum_{n,m} \left( M \dot u_{n,m}^2+M \dot v_{n,m}^2+J\dot \theta_{n,m}^2\right) \\&+\frac{1}{2} K_\theta\sum_{n,m} \Big[\left(\theta_{n,m} + \theta_{n-1,m}+2\theta_0\right)^2+(\theta_{n,m}+ \theta_{n,m-1}+2\theta_0)^2\Big] \\
	&+\frac{1}{2} K_l\sum_{n,m}\Big[ u_{n,m}-u_{n-1,m} -L\cos(\theta_{n,m}+\theta_0)-L\cos(\theta_{n-1,m}+\theta_0)+2L\cos\theta_0\Big]^2\\
 &+\frac{1}{2} K_l\sum_{n,m}\Big[ v_{n,m}-v_{n,m-1} -L\cos(\theta_{n,m}+\theta_0)-L\cos(\theta_{n,m-1}+\theta_0)+2L\cos\theta_0\Big]^2\\
&+\frac{1}{2} K_s\sum_{n,m}\Big[ u_{n,m}-u_{n,m-1} -(-1)^{n+m}L\sin(\theta_{n,m}+\theta_0)-(-1)^{n+m-1}L\sin(\theta_{n,m-1}+\theta_0)\Big]^2\\
 &+\frac{1}{2} K_s\sum_{n,m}\Big[ v_{n,m}-v_{n-1,m} +(-1)^{n+m}L\sin(\theta_{n,m}+\theta_0)+(-1)^{n+m-1}L\sin(\theta_{n-1,m}+\theta_0)\Big]^2,
	\end{split}
dddd	\end{align}
 where $L=\frac{a}{2\cos \theta_0}$ is half of the diagonal length of the square. Then, Hamilton's equations read
\begin{align}
&\begin{aligned}
M\ddot{u}_{n,m}&=-\frac{\partial H}{\partial u_{n,m}}
\end{aligned},\\
&\begin{aligned}
M\ddot{v}_{n,m}&=-\frac{\partial H}{\partial v_{n,m}}
\end{aligned},\\
&\begin{aligned}\label{Hamilton_eq3}
J\ddot{\theta}_{n,m}&=-\frac{\partial H}{\partial \theta_{n,m}}.
\end{aligned}
\end{align}

From Eq.~\ref{Hamiltonian} to Eq.~\ref{Hamilton_eq3}, the equations of motion (EOMs) for the square at site $(n,m)$ 
 can be derived as
	\begin{align}\label{governing_eqns_u}
		\begin{split}
		M \ddot{u}_{n,m} &= K_l \Big[u_{n+1,m}-u_{n,m}-L\cos(\theta_{n+1,m}+\theta_0)-L\cos(\theta_{n,m}+\theta_0)+2L
		\cos\theta_0\Big] \\
		&- K_l \Big[u_{n,m}-u_{n-1,m}-L\cos(\theta_{n,m}+\theta_0)-L\cos(\theta_{n-1,m}+\theta_0)+2L
		\cos\theta_0\Big]\\
		&+ K_s\Big[u_{n,m+1}-u_{n,m}-(-1)^{n+m+1}L\sin(\theta_{n,m+1}+\theta_0)-(-1)^{n+m}L\sin(\theta_{n,m}+\theta_0)\Big]\\
		&- K_s\Big[u_{n,m}-u_{n,m-1}-(-1)^{n+m}L\sin(\theta_{n,m}-\theta_0)-(-1)^{n+m-1}L\sin(\theta_{n,m-1}+\theta_0)\Big]
		\end{split}
		\end{align}
		
\begin{align}\label{governing_eqns_v}
\begin{split}
M \ddot{v}_{n,m} &= K_l \Big[v_{n,m+1}-v_{n,m}-L\cos(\theta_{n,m+1}+\theta_0)-L\cos(\theta_{n,m}+\theta_0)+2L
\cos\theta_0\Big]\\
&-K_l \Big[v_{n,m}-v_{n,m-1}-L\cos(\theta_{n,m}+\theta_0)-L\cos(\theta_{n,m-1}+\theta_0)+2L
\cos\theta_0\Big]\\
&+ K_s\Big[v_{n+1,m}-v_{n,m}+(-1)^{n+m+1}L\sin(\theta_{n+1,m}+\theta_0)+(-1)^{n+m}L\sin(\theta_{n,m}+\theta_0)\Big]\\
&- K_s\Big[v_{n,m}-v_{n-1,m}+(-1)^{n+m}L\sin(\theta_{n,m}+\theta_0)+(-1)^{n+m-1}L\sin(\theta_{n-1,m}+\theta_0)\Big]
\end{split}
\end{align}

\begin{align}\label{governing_eqns_theta}
\begin{split}
J \ddot{\theta}_{n,m} &= -K_{\theta} (\theta_{n-1,m} +\theta_{n+1,m}+\theta_{n,m+1} +\theta_{n,m-1}+4\theta_{n,m}+8\theta_0)\\ &-K_lL\sin(\theta_{n,m}+\theta_0)\Big[u_{n+1,m}-u_{n,m}-L\cos(\theta_{n+1,m}+\theta_0)-L\cos(\theta_{n,m}+\theta_0)+2L\cos\theta_0\Big]\\
&-K_lL\sin(\theta_{n,m}+\theta_0)\Big[u_{n,m}-u_{n-1,m}-L\cos(\theta_{n,m}+\theta_0)-L\cos(\theta_{n-1,m}+\theta_0)+2L\cos\theta_0\Big]\\
&-(-1)^{n+m-1}K_sL\cos(\theta_{n,m}+\theta_0)\Big[u_{n,m+1}-u_{n,m}+(-1)^{n+m}L\sin(\theta_{n,m+1}+\theta_0)\\&+(-1)^{n+m-1}L\sin(\theta_{n,m}+\theta_0)\Big]-(-1)^{n+m-1}K_sL\cos(\theta_{n,m}+\theta_0)\Big[u_{n,m}-u_{n,m-1}\\&+(-1)^{n+m-1}L\sin(\theta_{n,m}+\theta_0)+(-1)^{n+m-2}L\sin(\theta_{n,m-1}+\theta_0)\Big]\\
&-K_lL\sin(\theta_{n,m}+\theta_0)\Big[v_{n,m+1}-v_{n,m}-L\cos(\theta_{n,m+1}+\theta_0)-L\cos(\theta_{n,m}+\theta_0)+2L\cos\theta_0\Big]\\
&-K_lL\sin(\theta_{n,m}+\theta_0)\Big[v_{n,m}-v_{n,m-1}-L\cos(\theta_{n,m}+\theta_0)-L\cos(\theta_{n,m-1}+\theta_0)+2L\cos\theta_0\Big]\\
&+(-1)^{n+m-1}K_sL\cos(\theta_{n,m}+\theta_0)\Big[v_{n+1,m}-v_{n,m}-(-1)^{n+m}L\sin(\theta_{n+1,m}+\theta_0)\\&-(-1)^{n+m-1}L\sin(\theta_{n,m}+\theta_0)]+(-1)^{n+m-1}K_sL\cos(\theta_{n,m}+\theta_0)\Big[v_{n,m}-v_{n-1,m}\\&-(-1)^{n+m-1}L\sin(\theta_{n,m}+\theta_0)-(-1)^{n+m-2}L\sin(\theta_{n-1,m}+\theta_0)\Big]\\
&-T_{Morse}(\Delta{\theta_{n+1,m}})-T_{Morse}(\Delta{\theta_{n-1,m}})-T_{Morse}(\Delta{\theta_{n,m+1}})-T_{Morse}(\Delta{\theta_{n,m-1}})
\end{split}
\end{align}
where
\begin{align}\label{Tmorse}
\begin{split}
T_{Morse}(\Delta{\theta}) &= 2\alpha A\left[e^{2\alpha(\Delta\theta+2\theta_0-2\theta_{Morse})}-e^{\alpha(\Delta\theta+2\theta_0-2\theta_{Morse})}\right]\\
&-2\alpha A\Big[e^{-2\alpha(\Delta\theta+2\theta_0+2\theta_{Morse})}-e^{-\alpha(\Delta\theta+2\theta_0+2\theta_{Morse})}\Big],
\end{split}
\end{align}
and $\Delta{\theta_{n\pm1,m\pm1}}=\theta_{n,m}+\theta_{n\pm1,m\pm1}$. 

It is worth noting that Eqs.~\ref{governing_eqns_v}-\ref{Tmorse} are derived for a square connected to its four neighbors by thin ligaments. For certain squares at the boundary of the building blocks, some terms on the right-hand sides of Eqs.~\ref{governing_eqns_v}-\ref{Tmorse} are vanished due to the absence of ligaments. 

\section{Linear wave analysis}
In this section, we rely on the linearized EOMs to investigate the propagation of linear waves in a 2D periodic system of squares, in which each square is connected at its four vertices by thin ligaments. 

To linearize the equations around the initial equilibrium (i.e., when $\theta_0$ is infinitesimal), we employ the following approximations:
\begin{align}\label{Linearization1}
\begin{split}
\sin(\theta+\theta_0)&\approx \sin \theta_0 +\theta \cos\theta_0,\\
\cos(\theta+\theta_0)&\approx \cos \theta_0 -\theta \sin\theta_0,\\
T_{Morse}(\Delta{\theta})&\approx T_{Morse}(0)+T'_{Morse}(0)\Delta{\theta},
\end{split}
\end{align}
where
\begin{align}\label{Tmorse_0}
\begin{split}
T_{Morse}(0) &= 2\alpha A\Big[e^{2\alpha(2\theta_{0}-2\theta_{Morse})}-e^{\alpha(2\theta_{0}-2\theta_{Morse})}\Big]\\
&-2\alpha A\Big[e^{-2\alpha(2\theta_{0}+2\theta_{Morse})}-e^{-\alpha(2\theta_{0}+2\theta_{Morse})}\Big]
\end{split}
\end{align}
and
\begin{align}\label{Tmorse_d}
\begin{split}
T'_{Morse}(0) &= 2\alpha^2 A\Big[2e^{2\alpha(2\theta_{0}-2\theta_{Morse})}-e^{\alpha(2\theta_{0}-2\theta_{Morse})}\Big]\\
&+2\alpha^2 A\Big[2e^{-2\alpha(2\theta_{0}+2\theta_{Morse})}-e^{-\alpha(2\theta_{0}+2\theta_{Morse})}\Big]
\end{split}
\end{align}

Then, the linearized EOMs can be derived as

\begin{align}\label{governing_eqns_u_linear}
\begin{split}
M \ddot{u}_{n,m} &= K_l (u_{n-1,m} +u_{n+1,m}-2u_{n,m}) + K_s(u_{n,m-1} +u_{n,m+1}-2u_{n,m})\\
&+K_lL\sin\theta_0(\theta_{n+1,m}-\theta_{n-1,m})+(-1)^{n+m}K_sL\cos\theta_0 (\theta_{n,m+1}- \theta_{n,m-1}),
\end{split}
\end{align}

\begin{align}\label{governing_eqns_v_linear}
\begin{split}
M \ddot{v}_{n,m} &= K_s (v_{n-1,m} +v_{n+1,m}-2v_{n,m}) + K_l(v_{n,m-1} +v_{v,m+1}-2v_{n,m})\\
&+K_lL\sin\theta_0 (\theta_{n,m+1}-\theta_{n,m-1})+(-1)^{n+m}K_sL\cos\theta_0(-\theta_{n+1,m}+\theta_{n-1,m}),
\end{split}
\end{align}

\begin{align}\label{governing_eqns_theta_linear}
\begin{split}
J \ddot{\theta}_{n,m} &= -K_{\theta} (\theta_{n-1,m} +\theta_{n+1,m}+\theta_{n,m+1} +\theta_{n,m-1}+4\theta_{n,m}+8\theta_0)\\ &-8K_lL^2\cos^2\theta_0\theta_{n,m}-K_lL\sin\theta_0\Big[u_{n+1,m}+v_{n,m+1}-u_{n-1,m}-v_{n,m-1}\\&+L(\theta_{n+1,m}
+4\theta_{n,m}+\theta_{n-1,m}+\theta_{n,m+1}+\theta_{n,m-1})\sin\theta_0\Big]\\
&-(-1)^{n+m-1}K_sL\cos\theta_0(u_{n,m+1}+v_{n-1,m}-u_{n,m-1}-v_{n+1,m})\\ &+K_sL^2\cos^2\theta_0(\theta_{n,m+1}+\theta_{n,m-1}+\theta_{n+1,m}+\theta_{n-1,m}-4\theta_{n,m})\\
&-4T_{Morse}(0)-T'_{Morse}(0)(\theta_{n+1,m}+\theta_{n,m})-T'_{Morse}(0)(\theta_{n-1,m}+\theta_{n,m})\\
&-T'_{Morse}(0)(\theta_{n,m+1}+\theta_{n,m})-T'_{Morse}(0)(\theta_{n,m-1}+\theta_{n,m}).
\end{split}
\end{align}

A plane-wave solution of Eqs.~\ref{governing_eqns_u_linear}-\ref{governing_eqns_theta_linear}, with wave vector $\mathbf{k}$ and frequency $\omega$, can be expressed as 
\begin{align}\label{plane_wave_sol}
\mathbf{u}_{n,m}=\begin{bmatrix} u_{n,m} \\ v_{n,m} \\ \theta_{n,m} \end{bmatrix}=U\begin{bmatrix} \phi_u \\ \phi_v \\ \phi_\theta \end{bmatrix} e^{i(\mathbf{k}\cdot \mathbf{r}_{n,m}- \omega t)},
\end{align}
where $U$ is a constant, $\mathbf{r}_{n,m}$ is the position vector for square $(n,m)$, and $\boldsymbol{\phi}=\begin{bmatrix} \phi_u \\ \phi_v \\ \phi_\theta \end{bmatrix}$ is a modal vector. According to Floquet-Bloch theorem, the relations between displacements at neighboring sites are given as
\begin{align}\label{Bloch_condition}
\mathbf{u}_{n\pm1,m\pm1}=\mathbf{u}_{n,m}e^{i(\pm \xi_1 \pm \xi_2)},
\end{align}
where $\xi_1$ and $\xi_2$ are normalized wavenumbers along the two lattice vectors $\mathbf{e}_1$ and $\mathbf{e}_2$, respectively.

Substituting Eq.~\ref{plane_wave_sol} and Eq.~\ref{Bloch_condition} into Eqs.~\ref{governing_eqns_u_linear}-\ref{governing_eqns_theta_linear} yields the following eigenvalue problem 
\begin{align}\label{Eigenvalue_prob}
-\omega^2\mathbf{M}\mathbf{u}_n+\mathbf{K}\mathbf{u}_n=\mathbf{0},
\end{align}
where $\mathbf{M}=\begin{bmatrix} M & 0 & 0 \\ 0 & M & 0 \\ 0 & 0 & J \end{bmatrix}$ is a mass matrix, and $\mathbf{K}=\begin{bmatrix} K_{11} & K_{12} & K_{13} \\ K_{21} & K_{22} & K_{23} \\ K_{31} & K_{32} & K_{33} \end{bmatrix}$  is a stiffness matrix. All the components in $\mathbf{K}$ are given below:
\begin{align}\label{K_comp}
\begin{split}
K_{11}&= K_l(2-e^{-i\xi_1}-e^{i\xi_1})+K_s(2-e^{-i\xi_2}-e^{i\xi_2}),\\
K_{12}&=0,\\
K_{13}&=-K_lL\sin \theta_0 (e^{i\xi_1}-e^{-i\xi_1})-(-1)^{n+m}K_sL\cos\theta_0(e^{i\xi_2}-e^{-i\xi_2}),\\
K_{21}&=0,\\
K_{22}&=-K_s(e^{-i\xi_1}+e^{i\xi_1}-2)-K_l(e^{-i\xi_2}+e^{i\xi_2}-2),\\
K_{23}&=-K_lL\sin \theta_0 (e^{i\xi_2}-e^{-i\xi_2})-(-1)^{n+m}K_sL\cos\theta_0(e^{-i\xi_1}-e^{i\xi_1}),\\
K_{31}&=K_lL\sin\theta_0(e^{i\xi_1}-e^{-i\xi_1})-(-1)^{n+m}K_sL\cos\theta_0(e^{i\xi_2}-e^{-i\xi_2}),\\
K_{32}&=K_lL\sin\theta_0(e^{i\xi_2}-e^{-i\xi_2})-(-1)^{n+m}K_sL\cos\theta_0(e^{-i\xi_1}-e^{i\xi_1}),\\
K_{33}&=\Big[K_\theta+T'_{Morse}(0)+K_lL^2\sin^2\theta_0\Big] (4+e^{i\xi_1}+e^{-i\xi_1}+e^{i\xi_2}+e^{-i\xi_2})\\
&+K_sL^2\cos^2\theta_0(4-e^{i\xi_2}-e^{-i\xi_2}-e^{i\xi_1}-e^{-i\xi_1}).
\end{split}
\end{align}

The dispersion relation of the system can be obtained by solving the eigenvalue problem (i.e., Eq.~\ref{Eigenvalue_prob}). The system has three stable phases, each of which is associated with an equilibrium angle. By choosing the equilibrium angle $\theta_0=0,\,\pm \theta_1$, we can obtain the dispersion relations for all three phases, as displayed in Fig.~\ref{Bloch_analysis_SI}. Due to symmetry of the system and the multistable energy landscape, the two closed phases associated with $\pm \theta_1$ have identical dispersion relation.

\begin{figure*}[htbp]
    \centerline{ \includegraphics[width=1\textwidth]{Figures/Bloch_analysis_SI.pdf}}
    \caption{Dispersion relations for (a) the open phase ($\theta_0=0$) and (b) two closed phases ($\theta_0=\pm\theta_1$).} 
    \label{Bloch_analysis_SI}
\end{figure*}

\section{Steady-state analysis}

To investigate the linear dynamic behavior of the finite-sized multistable metamaterial, we  perform a steady-state analysis to obtain the frequency response. We start with the EOMs of the system:
\begin{align}\label{Steady-state eqn0}
\mathbf{M}\ddot{\mathbf{u}}+\mathbf{K}\mathbf{u}=\mathbf{F},
\end{align}
in which $\mathbf{M}$, $\mathbf{K}$, $\mathbf{F}$, and $\mathbf{u}$ are the global mass matrix, stiffness matrix, external force, and displacemtn vector, respectively. Here, the stiffness matrix $\mathbf{K}$ depends on the current configuration of the system, i.e., $\mathbf{K}(\bar{\mathbf{u}})$ in which $\bar{\mathbf{u}}$ is the current displacement vector. Since we are interested in harmonic excitations, the external force $\mathbf{F}$ can be written in the form 
\begin{align}\label{External force}
\mathbf{F}=\mathbf{F}_0 e^{i\omega t},
\end{align}
where $\mathbf{F}_0 $ is a constant amplitude vector. The steady-state response can be expressed as
\begin{align}\label{Disp u harmonic}
\mathbf{u}=\mathbf{U}(i\omega) e^{i\omega t},
\end{align}
where $\mathbf{U}(i\omega)$ is a complex function of the driving frequency $\omega$ and the parameters of the system. Substituting Eq.~\ref{External force} and Eq.~\ref{Disp u harmonic} into Eq.~\ref{Steady-state eqn0} gives 

\begin{align}\label{Steady-state eqn}
(-\omega^2\mathbf{M}+\mathbf{K})\mathbf{U}(i\omega)=\mathbf{F}_0,
\end{align}

To obtain the global stiffness $\mathbf{K}$, we first derive the element stiffness matrix $\mathbf{K}_e$
\begin{align}\label{Steady-state eqn}\mathbf{K}_e=\begin{bmatrix} K_{11} & K_{12} & K_{13} & K_{14} & K_{15} & K_{16}\\ K_{21} & K_{22} & K_{23} & K_{24} & K_{25} & K_{26}\\ K_{31} & K_{32} & K_{33} & K_{34} & K_{35} & K_{36}\\K_{41} & K_{42} & K_{43} & K_{44} & K_{45} & K_{46}\\ K_{51} & K_{52} & K_{53} & K_{54} & K_{55} & K_{56}\\ K_{61} & K_{62} & K_{63} & K_{64} & K_{65} & K_{66}\end{bmatrix}
\end{align}
For an element consisting of two squares $(n,m)$ and $(n+1,m)$ that are located horizontally, the components  can be derived as 
\begin{align}\label{K_comp_non_h}
\begin{split}
K_{11}&=K_l, K_{12}=0, K_{13}=-K_lL\sin\bar{\theta}_{n,m}, K_{14}=-K_l,K_{15}=0, K_{16}=-K_lL\sin\bar{\theta}_{n+1,m},\\
K_{21}&=0, K_{22}=K_s, K_{23}=(-1)^{n_1}K_sL\cos\bar{\theta}_{n,m}, K_{24}=0,K_{25}=-K_s,\\ K_{26}&=(-1)^{n_2}K_sL\cos\bar{\theta}_{n+1,m},
K_{31}=K_{13}, K_{32}=K_{23},\\
K_{33}&= K_\theta+K_lL^2\sin^2\bar{\theta}_{n,m}+K_sL^2\cos^2\bar{\theta}_{n,m}+T'_{Morse}(\Delta\bar{\theta})\\&+K_lL\cos\bar{\theta}_{n,m}\Big[\bar{u}_{n+1,m}-\bar{u}_{n,m}-L\cos\bar{\theta}_{n+1,m}-L\cos\bar{\theta}_{n,m}+2L\cos\theta_0\Big]\\
&+(-1)^{n_1}K_sL\sin\bar{\theta}_{n,m}\Big[\bar{v}_{n+1,m}-\bar{v}_{n,m}-(-1)^{n_2}L\sin\bar{\theta}_{n+1,m}-(-1)^{n_1}L\sin\bar{\theta}_{n,m}\Big],\\
K_{34}&=K_lL\sin\bar{\theta}_{n,m}, K_{35}=-(-1)^{n_1}K_sL\cos\bar{\theta}_{n,m},\\
K_{36}&=K_\theta+K_lL^2\sin\bar{\theta}_{n,m}\sin\bar{\theta}_{n+1,m}-K_sL^2\cos\bar{\theta}_{n,m}\cos\bar{\theta}_{n+1,m}+T'_{Morse}(\Delta\bar{\theta})\\
K_{41}&=K_{14}, K_{42}=K_{24}, K_{43}=K_{34},K_{44}=K_l,K_{45}=0, K_{46}=K_lL\sin\bar{\theta}_{n+1,m},\\
K_{51}&=K_{15}, K_{52}=K_{25}, K_{53}=K_{35},K_{54}=K_{45},K_{55}=K_s, K_{56}=-(-1)^{n_2}K_sL\cos\bar{\theta}_{n+1,m},\\K_{61}&=K_{16}, K_{62}=K_{26}, K_{63}=K_{36},K_{64}=K_{46},K_{65}=K_{56},\\
K_{66}&=K_\theta+K_lL^2\sin^2\bar{\theta}_{n+1,m}+K_sL^2\cos^2\bar{\theta}_{n+1,m}+T'_{Morse}(\Delta\bar{\theta})\\
&+K_lL\cos\bar{\theta}_{n+1,m}\Big[\bar{u}_{n+1,m}-\bar{u}_{n,m}-L\cos\bar{\theta}_{n+1,m}-L\cos(\theta_1)+2L\cos\theta_0\Big]\\
&+(-1)^{n_2}K_sL\sin\bar{\theta}_{n+1,m}\Big[\bar{v}_{n+1,m}-\bar{v}_{n,m}-(-1)^{n_2}L\sin\bar{\theta}_{n+1,m}-(-1)^{n_1}L\sin\bar{\theta}_{n,m}\Big],
\end{split}
\end{align}
where $n_1=n+m$, $n_2=n+m+1$, and 
$\Delta\bar{\theta}=\bar{\theta}_{n,m}+\bar{\theta}_{n+1,m}$ (note that $\bar{(\cdot})$ denotes the current displacement). For an element consisting of two squares $(n,m)$ and $(n,m+1)$ that are located vertically, the components of $\mathbf{K}_e$ become 
\begin{align}\label{K_comp_non_v}
\begin{split}
K_{11}&=K_s, K_{12}=0, K_{13}=-(-1)^{n_1}K_sL\cos\bar{\theta}_{n,m}, K_{14}=-K_s,K_{15}=0, K_{16}=-(-1)^{n_2}K_sL\cos\bar{\theta}_{n,m+1},\\
K_{21}&=0, K_{22}=K_l, K_{23}=-K_lL\sin\bar{\theta}_{n,m}, K_{24}=0,K_{25}=-K_l,\\ K_{26}&=-K_lL\sin\bar{\theta}_{n,m+1},
K_{31}=K_{13}, K_{32}=K_{23},\\
K_{33}&= K_\theta+K_lL^2\sin^2\bar{\theta}_{n,m}+K_sL^2\cos^2\bar{\theta}_{n,m}+T'_{Morse}(\Delta\bar{\theta})\\&+K_lL\cos\bar{\theta}_{n,m}\Big[\bar{v}_{n,m+1}-\bar{v}_{n,m}-L\cos\bar{\theta}_{n,m+1}-L\cos\bar{\theta}_{n,m}+2L\cos\theta_0\Big]\\
&-(-1)^{n_1}K_sL\sin\bar{\theta}_{n,m}\Big[\bar{u}_{n,m+1}-\bar{u}_{n,m}+(-1)^{n_2}L\sin\bar{\theta}_{n,m+1}+(-1)^{n_1}L\sin\bar{\theta}_{n,m}\Big],\\
K_{34}&=(-1)^{n_1}K_sL\cos\bar{\theta}_{n,m}, K_{35}=K_lL\sin\bar{\theta}_{n,m},\\
K_{36}&=K_\theta+K_lL^2\sin\bar{\theta}_{n,m}\sin\bar{\theta}_{n,m+1}-K_sL^2\cos\bar{\theta}_{n,m}\cos\bar{\theta}_{n,m+1}+T'_{Morse}(\Delta\bar{\theta})\\
K_{41}&=K_{14}, K_{42}=K_{24}, K_{43}=K_{34},K_{44}=K_s,K_{45}=0, K_{46}=(-1)^{n_2}K_sL\cos\bar{\theta}_{n,m+1},\\
K_{51}&=K_{15}, K_{52}=K_{25}, K_{53}=K_{35},K_{54}=K_{45},K_{55}=K_l, K_{56}=K_lL\sin\bar{\theta}_{n,m+1},\\K_{61}&=K_{16}, K_{62}=K_{26}, K_{63}=K_{36},K_{64}=K_{46},K_{65}=K_{56},\\
K_{66}&=K_\theta+K_lL^2\sin^2\bar{\theta}_{n,m+1}+K_sL^2\cos^2\bar{\theta}_{n,m+1}+T'_{Morse}(\Delta\bar{\theta})\\
&+K_lL\cos\bar{\theta}_{n,m+1}\Big[\bar{v}_{n,m+1}-\bar{v}_{n,m}-L\cos\bar{\theta}_{n,m+1}-L\cos(\theta_1)+2L\cos\theta_0\Big]\\
&-(-1)^{n_2}K_sL\sin\bar{\theta}_{n,m+1}\Big[\bar{u}_{n,m+1}-\bar{u}_{n,m}+(-1)^{n_2}L\sin\bar{\theta}_{n,m+1}+(-1)^{n_1}L\sin\bar{\theta}_{n,m}\Big].
\end{split}
\end{align}

Then, the global stiffness matrix $\mathbf{K}$ can be obtained by assembling the element stiffness matrix $\mathbf{K}_e$. For a given configuration $\bar{\mathbf{u}}$, we can obtain the steady-state response of the multistable system for harmonic excitations by solving Eq.~\ref{Steady-state eqn0}. As shown in Fig.~4A-B in the main text, the steady-state responses for the all-open and all-closed configurations are highly consistent with the corresponding dispersion relations for the open and closed phases (i.e., Fig.~\ref{Bloch_analysis_SI}). This indicates that the linear wave analysis presented above can be employed as useful guidelines for the design of the multistable system with desired dynamic properties.